# Phononic thermal properties of two-dimensional materials


Xiaokun Gu[1,2], Yujie Wei,[3*] Xiaobo Yin[2,4], Baowen Li[2*] and Ronggui Yang[2,4*]

[1]Institute of Engineering Thermophysics, Shanghai Jiao Tong University, Shanghai, China 200240

[2]Department of Mechanical Engineering, University of Colorado, Boulder, CO, USA 80309

[3]LNM, Institute of Mechanics, Chinese Academy of Sciences, Beijing, China, 100190

[4]Materials Science and Engineering Program, University of Colorado, Boulder, CO, USA 80309

*Emails: Ronggui.Yang@Colorado.Edu;

Baowen.Li@Colorado.Edu;

Yujie.Wei@lnm.imech.ac.cn


Following the emergence of many novel two-dimensional (2-D) materials beyond graphene, interest has grown in exploring implications for fundamental physics and practical applications, ranging from electronics, photonics, phononics, to thermal management and energy storage. In this Colloquium, we first summarize and compare the phonon properties, such as phonon dispersion and relaxation time, of pristine 2-D materials with single layer graphene to understand the role of crystal structure and dimension on thermal conductivity. We then compare the phonon properties, contrasting idealized 2-D crystals, realistic 2-D crystals, and 3-D crystals, and synthesizing this to develop a physical picture of how the sample size of 2-D materials affects their thermal conductivity. The effects of geometry, such as number of layers, and nanoribbon width, together with the presence of defects, mechanical strain, and substrate interactions, on the thermal properties of 2-D materials are discussed. Intercalation affects both the group velocities and phonon relaxation times of layered crystals and thus tunes the thermal conductivity along



both the through-plane and basal-plane directions. We conclude with a discussion of the challenges in theoretical and experimental studies of thermal transport in 2-D materials. The rich and special phonon physics in 2-D materials make them promising candidates for exploring novel phenomena such as topological phonon effects and applications such as phononic quantum devices.

## Table of Contents





# I. Introduction

A thorough understanding of heat conduction, on one hand, is of primary importance for fundamental research such as constructing microscopic pictures for macroscopic irreversible heat transfer; and, on the other hand, provides theoretical basis for thermal energy control and management. The latter is crucial in our daily life, ranging from thermal management of electronic devices like smartphones and power inverters in hybrid or all electric vehicles to power generation and energy utilization processes. In the macroscopic world, thermal conductivity is used as an indicator for heat conducting capability of a material, implying that the phenomenological Fourier's law of heat conduction, which states that the heat flux is proportional to the temperature gradient, is valid.

With the rapid development of novel materials, in particular low-dimensional materials, like quasi-one dimensional (1-D) nanowires, nanotubes, polymer fibers, and two-dimensional (2-D) materials, like graphene and transition metal dichalcogenides, one may ask: is Fourier's law of heat conduction (which describes diffusive transport of energy carriers) still valid in such low-dimensional materials? Answering this question is not trivial: it not only enriches our understanding of the nature of heat conduction but also may provide theoretical guidelines for applications of low-dimensional materials.

In the last few decades, we have witnessed important progress in this direction, resulted from the great efforts made by physicists, mathematicians, and engineers. First of all, important progress in understanding of phonon transport in (quasi) 1-D systems has been achieved. Through many intensive numerical simulations (Lepri et al. 2003, Dhar 2008, Lepri et al. 2016) and mathematical proofs (Prosen and Campbell 2000), thermal conductivity is found to not be an intensive property. Rather, thermal conductivity is divergent with system size in many 1D



momentum conserved lattices and toy models (more detailed discussions are given in Sato 2016). Different approaches such as hydrodynamics and renormalization group theory (Narayan and Ramaswamy, 2002;  Mai and Narayan, 2006), mode coupling theory (Lepri, 1998;  Lepri, *et al.*, 1998), and Peierls-Boltzmann transport theory (Pereverzev, 2003) have been employed to show that the thermal conductivity follows $K \sim L^z$, where $L$ is the sample length and the exponent $z$ varies from system to system. This divergent thermal conductivity is shown to be connected with super diffusion of phonons (Liu et al. 2014).

The phononic thermal conduction in quasi 1-D system might not be the same as that in 1-D lattice and toy models, because the atoms in the quasi 1-D nanostructures can vibrate in the three-dimensional (3-D) real space. However, anomalous thermal conduction behavior has also been observed in numerical simulations for carbon nanotubes (Maruyama, 2002;  Zhang and Li, 2005), nanowires (Yang, *et al.*, 2010), and polymer chains (Henry and Chen, 2008;  2009;  Liu and Yang, 2012). It has been pointed out (Mingo and Broido, 2005) that if only the first-order three-phonon scatterings are considered, the thermal conductivity of nanotubes is indeed divergent with length, where the exponent $z$ has a value between 1/3 and 1/2. However, if the three-phonon scattering to the second order is included in the calculations, the thermal conductivity converges to a finite value, but for very long nanotubes.  This theoretical prediction has not yet been experimentally confirmed.

These theoretical and numerical studies have been naturally extended to 2-D systems, although the studies are much more challenging as the degrees of freedom are significantly increased. Both the mode coupling theory (Ernst, *et al.*, 1971;  1976a;  1976b;  Lepri, *et al.*, 2003), renormalization theory (Narayan and Ramaswamy, 2002), and numerical simulations (Lippi and Livi, 2000;  Yang, *et al.*, 2006;  Xiong, *et al.*, 2010;  Wang, *et al.*, 2012a) show that



thermal conductivity in 2-D anharmonic lattices diverges logarithmically with system size ($L \times L$), namely $\kappa \sim \log(L)$. However, a rigorous mathematical proof from fundamental symmetry of the system and fundamental law of conservation like in 1-D systems (Prosen and Campbell 2000), is still lacking for the 2-D systems.

On the other hand, the novel 2-D materials exemplified by graphene provide a great playground for scientists and engineers. In particular, the pioneering experimental measurements using optothermal Raman method (Balandin, *et al.*, 2008) showed that single-layer graphene may have a thermal conductivity between 2000-5000 W/(m K) at room temperature, even higher than that of diamond. This result has inspired many studies aiming at both understanding its underlying physics and exploring potential applications for heat exchangers and thermal management. Notably, those studies on phonon transport and thermal conductivity of graphene have laid a solid foundation for other emerging 2-D materials, as reviewed in (Balandin 2010; Nika and Balandin, 2012; Nika and Balandin, 2017; Pop, *et al.*, 2012; Sadeghi, *et al.*, 2012; Wang, *et al.*, 2014; Xu, *et al.*, 2014, Li et al, 2017).

Over the past decade, many novel 2-D materials have emerged. Some of them have been shown to have similar or even superior electronic and optical properties than graphene, and are proposed for a wide range of applications, such as field-effect transistors, light-emitting diodes, spin- and valley-tronics, topological insulators, catalysts, and energy storage. Understanding phononic thermal properties of such 2-D materials could be very important for the design of novel 2-D materials-enabled applications. However, most of these 2-D materials are different from the "one-atom-thick" single-layer graphene, making the direct application of existing knowledge of graphene to other 2-D materials questionable.



Controlling the thermal properties of 2-D materials could enable new generations of thermoelectric materials, thermal insulators, and even phononic computing devices. The best thermoelectric materials, such as $Bi_2Te_3$-based alloys, are indeed layered materials. The diversity of 2-D materials could lead to novel thermoelectric materials and fully-dense superior thermal insulators with ultralow thermal conductivity by manipulating phonon mismatch between different layers of 2-D materials. 2-D materials may offer a novel platform to realize phonon thermal diodes, phonon transistors, topological phonon insulators, and even quantum memories and quantum sensors.

With the rapid discovery of novel 2-D materials, and the increasing number of papers on the phonon and thermal properties of 2-D materials, it is desirable to have a comprehensive picture about the state-of-the-art of this field. Indeed, there are already quite a few recent review articles on distinct aspects of the thermal properties of 2-D materials in addition to the aforementioned reviews on graphene: Wang et al. (Wang*, et al.*, 2017) summarize the progress of measurement methods, and Xu et al. (Xu*, et al.*, 2016) provide a critical review on the advantage and disadvantage of existing numerical and experimental techniques, and the respective challenging issues. While Zhang and Zhang (Zhang and Zhang, 2017) are more focused on the cutting-edge thermal devices such as thermal diodes and thermal modulators with graphene and $MO_2$, Qin and Hu (Qin and Hu, 2016) review the diversity of thermal properties in other 2-D materials.

In this Colloquium, we aim to provide a physical picture about how phonon scattering in 2-D materials influences their thermal transport properties. This Colloquium will not just summarize the existing results (including numerical, theoretical and experimental), but will give a pedagogical introduction to the physical picture of phonon scattering and thermal properties in



both pristine 2-D materials and their use in practical applications. The following fundamental questions will be answered:

a) Can phonons in a suspended single layer 2-D material undergo diffusive motions that lead to a normal heat conduction, *i.e.*, is the thermal conductivity independent of dimensions such as width and length?  Since suspended single layer 2-D materials are very fragile and not easy to manipulate, they might not be suitable for many practical applications. A quite logical step is to substitute them with multilayer thin films. How does the inter-layer interaction affect the phonon transport?

b) In most applications, 2-D materials sit on a substrate or are in contact with a metal electrode. How does the substrate or the electrode material affect phonon transport? This is an important question since the substrate induces additional channels for phonon transport and phonon scattering.

c) Imperfections, such as point defects, vacancies and grain boundaries, inevitably occur in 2-D materials. How do imperfections and mechanical strains change the thermal conductivity?

d) It is rather exciting that 2-D materials can be functionalized and intercalated. How can the thermal properties of layered 2-D materials be tuned through surface functionalization and intercalation?

We are to answer these aforementioned questions in Part II from the phonon scattering picture, aiming to give readers a comprehensive physical picture about thermal behavior of 2-D materials and its connection with the crystal structures of these materials. Discussions in Part II are focused on pristine 2-D materials, neglecting defect scattering, boundary scattering, and interface scattering. Since defects and impurities are ubiquitous in real materials, in Part III we



discuss how defects interfere with phonons and subsequently affect the thermal conductivity. More importantly, surface functionalization and chemical intercalation render great opportunity to tune the thermal conductivity of 2-D materials that could enable new functionalities. In Part IV, we summarize the main discoveries of current research on phonon and thermal properties of 2-D materials, along with the challenges and difficulties in both experimental and theoretical studies. We also suggest some potentially fruitful research directions for 2-D materials including electronic thermal transport, topological phonon effect, quantum memory and sensors.

## II.  Phonon and thermal properties of pristine 2-D materials

### A.  Basics of thermal transport in 2-D materials

The ability of a solid to conduct heat is characterized by thermal conductivity, $\kappa$, which relates a resultant heat flux $J$ to an applied temperature gradient $\nabla T$ via the phenomenological Fourier's law of heat conduction, $J = -\kappa \cdot \nabla T$. Depending on the crystal structures, the thermal conductivity can be a scalar in ideally isotropic materials or a tensor in anisotropic materials.

For semiconductors and insulators, heat is mainly carried by phonons, the quantized lattice vibration. According to the kinetic theory (Ziman, 1960; Ashcroft and Mermin, 1978; Chen, 2005), the thermal conductivity in a crystalline material is determined by mode-dependent heat capacity $c$, group velocity $v$ and relaxation time $\tau$ or mean free path $\Lambda$ through

$$k = \sum_{\mathbf{q}s} c_{\mathbf{q}s} v_{\mathbf{q}s}^2 t_{\mathbf{q}s} = \sum_{\mathbf{q}s} c_{\mathbf{q}s} v_{\mathbf{q}s} \mathsf{L}_{\mathbf{q}s} \tag{1},$$

where the subscript $\mathbf{q}s$ denotes a phonon mode in the $s$-th phonon branch with a momentum $\hbar \mathbf{q}$ ($\hbar$ is the Planck constant). The mode-dependent heat capacity $c_{\mathbf{q}s}$ is expressed as $\hbar \omega_{\mathbf{q}s} \partial f_{\mathbf{q}s}^0 / \partial T$, where $f_{\mathbf{q}s}^0 = 1/\left( \exp(\hbar \omega_{\mathbf{q}s} / k_B T) - 1 \right)$ is the equilibrium phonon population function obeying the



Bose-Einstein statistics, $\omega_{\mathbf{q}s}$ is the phonon frequency, and $k_{\mathrm{B}}T$ the product of the Boltzmann constant and the absolute temperature. The phonon dispersion and phonon scattering in this equation can be calculated to reasonably good accuracy by using the first-principles-based calculations (Esfarjani, *et al.*, 2011; Garg, *et al.*, 2011; Li, *et al.*, 2012b). The thermal conductivity can then be determined through the Peierls-Boltzmann transport equation (PBTE) (Broido, *et al.*, 2007; Ward, *et al.*, 2009; Li, *et al.*, 2012c; Fugallo, *et al.*, 2013).

In engineering applications, thermal conductivity is generally regarded as an intensive material property that is independent of sample size. This, however, is only valid when phonon transport is diffusive (Chen, 2001). When the sample size along the transport direction is much smaller than the phonon mean free path $\Lambda_{\mathbf{q}s}\left(= v_{\mathbf{q}s}\tau_{\mathbf{q}s}\right)$, phonons propagate ballistically across the sample without experiencing appreciable scattering. This is the case in thin films where the thermal conductivity decreases as films become thinner due to the increase of phonon-boundary scattering (Ju and Goodson, 1999). The diffusive phonon transport entails a constant thermal conductivity for sufficiently large samples of 3-D materials. However, for some low-dimensional systems, especially in 1-D systems (Lepri, *et al.*, 2003; Yang, *et al.*, 2012), the thermal conductivity does not necessarily converge to a constant value even when the length of a 1-D sample is large because there is no scattering mechanism to guarantee diffusive transport.

Even though no rigorous proof is available on whether heat conduction in 2-D materials obey Fourier's law of heat conduction, to be detailed in Sec. II.B, we adopt the terminology "effective" thermal conductivity used in most literatures and compare the heat conducting capability of a few 2-D materials, as shown in <span style="color:red">Fig. 1</span>. The temperature-dependent thermal conductivity values from both the first-principles calculations (Fugallo, *et al.*, 2014; Gu and Yang, 2014; Lindsay, *et al.*, 2014; Zhu, *et al.*, 2014; Cepellotti, *et al.*, 2015; Gu and Yang, 2015; Jain and



McGaughey, 2015; Wang, *et al.*, 2015; Peng, *et al.*, 2016; Qin, *et al.*, 2016a; Zeraati, *et al.*, 2016) and from experimental measurements (Cai, *et al.*, 2010; Chen, *et al.*, 2010; Faugeras, *et al.*, 2010; Lee, *et al.*, 2011; Chen, *et al.*, 2012; Yan, *et al.*, 2014; Peimyoo, *et al.*, 2015; Zhang, *et al.*, 2015b), when available, are presented. Experimental measurement of thermal conductivity of 2-D materials is still a challenging task. Currently, there are two main techniques to measure the basal-plane (in-plane) thermal conductivity of 2-D materials: opto-thermal Raman method (Balandin, *et al.*, 2008) and micro-bridge method (Kim, *et al.*, 2001, Shi, *et al.*, 2003). The thermal conductivity of 2-D materials with sample size around 10-μm-long or sufficiently large, spans about three orders of magnitude from tens of W/(m K) to thousands of W/(m K) at room temperature.

In this Colloquium, we try to correlate the values of thermal conductivity of 2-D materials with their crystal structures and atom species. This could not only provide a first-order quantitative estimate on the thermal conductivity value of emerging 2-D materials when exact calculations and measurements are not available, but also deepen our understanding on the diverse dependence of the thermal conductivity of many 2-D materials on various physical factors.

Indeed, the efforts to connect crystal structures and atom species with the macroscopic thermal conductivity of bulk materials can be dated back to 1970s. Among many models, the Slack equation, (Slack, 1973; Morelli and Heremans, 2002)

$$\kappa = A \frac{\bar{M} \theta_a^3 \delta n^{1/3}}{\gamma_G^2 T}, \qquad (2)$$

is the mostly used one. In this equation, $n$ is the number of atoms in the primitive unit cell, $\delta$ is the lattice size, $\theta_a$ is the Debye temperature of acoustic branches, $\bar{M}$ is the average mass of the



atoms in the crystal, $\gamma_G$ is the Grüneisen parameter, and $A$ is a constant that equals to $3.1 \times 10^6$ (with the unit of $\kappa$ in W/(m K), $\bar{M}$ in atomic mass unit, $T$ in Kelvin, and $\delta$ in Å). The Grüneisen parameter describes overall effect of volume change of a crystal on vibrational properties, which is expressed as $\gamma_G = V / C_V (dP / dT)_V$, where $V$ is the volume of a crystal, $C_V$ the heat capacity at constant volume, and $(dP / dT)_V$ pressure change due to temperature variation at constant volume. Since the Debye temperature decreases with atomic mass, and increases with bonding stiffness, crystals with lighter atomic mass and stronger bonds will have larger $\bar{M} \theta_a^3$ and thus a larger thermal conductivity. On the other hand, with the increase of $n$, the size of the first Brillouin zone decreases as the volume of primitive cell increases. Assuming the speed of sound is independent on $n$, the Debye temperature of the acoustic branches, which are roughly proportional to product of the size of the first Brillouin zone and the sound speed, becomes lower for larger $n$. As a result, a larger $n$ (number of atoms in the primitive unit cell) leads to a smaller thermal conductivity. Eq. (2) has indeed been used as a guideline in searching for high thermal conductivity materials where the requirements are: 1) high Debye temperature, 2) small atomic mass, 3) simple crystal structure, and 4) low anharmonicity. Although the Slack equation is derived for 3-D materials, these guidelines could be adopted to understand the thermal conductivity of 2-D materials. To evaluate the thermal conductivity by Eq. (2), we need to determine the Debye temperature $\theta_a$ and the Grüneisen parameter $\gamma_G$. Here, we first estimate the Debye temperature of the $i$-th acoustic branch, $\theta_{a,i}$ using the group velocity at $\Gamma$ point ($\mathbf{q} = 0$), $v_i$ through $\theta_{a,i} = \hbar v_i q_{cut} / k_B$, with the cutoff wave vector at the zone boundary. Then, $\theta_a$ is set as the averaged Debye temperature of the two in-plane acoustic branches,



$\theta_a = \left( \dfrac{1}{2} \displaystyle\sum_{i=1,2} \theta_{a,i}^2 \right)^{1/2}$ . The Grüneisen parameter $\gamma_G$ is chosen to be 2.0 here as an estimate for all 2-

D materials, which was used for graphite earlier by Klemens and Pedraza (Klemens and Pedraza,

1994), though more accurate mode-dependent $\gamma_G$ can be obtained through first-principles

calculations (Mounet and Marzari, 2005). In Fig. 2, we compare the first principles calculations

and the predictions using the Slack equation for the room temperature thermal conductivity of 2-

D materials. We see that the Slack equation prediction (Eq. (2)) matches the general trend for

thermal conductivity in these known 2-D materials. The success of the Slack equation is due to

its incorporation of a rough estimate of how all factors in (Eq. (1)) (i.e. the group velocity, the

heat capacity and the relaxation time) depend on the basic crystal properties. While there are

some outliers, such as graphene and WS$_2$, whose thermal conductivity exceeds the predictions of

the Slack equation, these deviations could be attributed to the assumption that the Grüneisen

constant $\gamma_G = 2.0$ for all the 2-D materials instead of varying between materials. In addition, the

in-plane modes and flexural out-of-plane modes are not treated independently in the Slack model

and the Debye temperature for the flexural out-of-plane acoustic branches is set to zero. As such,

the thermal conductivity contributions from the flexural out-of-plane branches are not correctly

incorporated in the Slack model.

To gain insights into differences between predictions from the Slack equation and that from

the first-principles calculations, we show in Fig. 3 the room temperature thermal conductivity

of a few 2-D materials, sorted in descending order. Interestingly, the high thermal conductivity

materials (such as graphene and boron nitride) have planar structures, the moderate thermal

conductivity materials (such as 2H transition metal dichalcogenides (TMDs)) have tri-layer,

mirror symmetric crystal structures, and the low thermal conductivity materials (such as 1T



TMDs and buckled materials) have bi- or tri-layer crystal structure without mirror symmetry. This observation reveals that the thermal conductivity of 2-D materials is highly sensitive to the crystal structure, in addition to the physical factors that are already included in the Slack equation.

To understand the observed correlation between the thermal conductivity and the crystal structure, we first recall the symmetry selection rule in planar 2-D materials (Lindsay, *et al.*, 2010b). If the 2-D material is planar, the out-of-plane motion of atoms is completely decoupled from the in-plane motion. As a result, it is possible to classify phonon modes into flexural out-of-plane modes including flexural acoustic (ZA) and flexural optical (ZO) modes, and the in-plane modes that including longitudinal acoustic (LA), longitudinal optical (LO), transverse acoustic (TA), and transverse optical (TO) modes. Meanwhile, the planar structure could make the third-order anharmonic force constants, $\psi_{i,j,k}^{\alpha\beta\gamma} = \partial^3 E / \partial u_i^\alpha \partial u_j^\beta \partial u_k^\gamma$, zero, (the third-order derivative of the total energy of the crystal, $E$, with respect to atom displacements including the displacement of atom $i$ ($j$ and $k$) along direction $\alpha$ ($\beta$ and $\gamma$), if one or three of the three directions $\alpha$, $\beta$, $\gamma$ is the two through-plane directions. This means that simultaneously displacing the three components corresponding to these zero force constants $\psi$ does not lead to (anharmonic) phonon-phonon scattering. When the three-phonon scatterings involving an odd number of flexural phonon modes, including acoustic (ZA) and optical (ZO) modes, one can easily realize that all the third-order force constants are zero by noticing that the atomic motions of flexural modes are purely along the through-plane $z$ direction and those of in-plane modes are purely in the basal plane. Therefore, these three-phonon scattering processes cannot happen (i.e. have a scattering rate of zero), leading to much larger relaxation times for the flexural acoustic phonon modes (Lindsay, *et al.*, 2010b). This unique selection rule for structures of planar symmetry is



one of the main reasons for the large thermal conductivity observed in both the single layer graphene and boron nitride (Lindsay*, et al.*, 2010b; 2011; Lindsay and Broido, 2012).

Since 2H TMDs are not planar crystals, the flexural phonon modes are no longer purely out-of-plane vibration. Therefore, the symmetry selection rule breaks down so that the three-phonon scattering processes involving odd number of flexural phonons are allowed. However, the tri-plane symmetric structure ensures $\psi_{i,j,k}^{\alpha\beta\gamma}$ to be always equal to zero if $i$, $j$ and $k$ denote metal atoms in the middle layer, i.e. that there is no anharmonicity induced by the relative motion of three metal atoms along $z$ direction. Since the dominant atomic motion of ZA phonons is along the $z$ direction, the scattering rates of these ZA phonon modes are thus very small (Gu*, et al.*, 2016). For other 2-D crystals, such as buckled honeycomb crystals (Gu and Yang, 2015) and 1T TMDs (Gu and Yang, 2014), there is no symmetry in their crystals which results in stronger phonon scattering compared to planar crystals and 2H TMDs.

Apart from the crystal structure, the atomic masses (atom number) also play an important role in explaining the disparity of thermal conductivity in the group of materials with identical crystal structure. For a few TMDs, the mass of metal atoms is much larger than the chalcogen atoms. Large mass difference of the basis atoms can lead to a large frequency gap between the optical and acoustic phonon branches (Gu and Yang, 2014). For example, the acoustic-optical (a-o) frequency gap in $WS_2$ is as large as 110 cm$^{-1}$, which is close to the range of acoustic phonons of $WS_2$ (0-178 cm$^{-1}$), while the gap is only 45 cm$^{-1}$ in $MoS_2$, much smaller than the frequency range of acoustic phonons (0-230 cm$^{-1}$). Because of the large phonon frequency gap in $WS_2$ and the energy conservation requirement of three-phonon scattering, an important phonon scattering channel (the annihilation process of two acoustic phonon modes into one optical one (acoustic + acoustic -> optical)) is strongly suppressed (but not prohibited). As a result, a weaker phonon-



phonon scattering rate is observed in $WS_2$, which yields a much higher thermal conductivity in $WS_2$ than in $MoS_2$. While the above discussion assumed that three-phonon scatterings were the dominant scattering mechanisms for phonons, recent advances in the theoretical studies have shown that four-phonon scatterings might be important for bulk semiconductors with large frequency gap, like boron arsenide (BAs), due to the much larger phonon scattering phase space of four-phonon scatterings compared to the three-phonon scatterings (Feng *et al.*, 2017). We note that the contribution of four-phonon scatterings to phonon transport in 2-D materials has not yet been studied. It would be desirable to include four-phonon scatterings, if possible, when studying the thermal transport of 2-D materials with large atomic mass difference.

## B. The validity of Fourier's law

While thermal conductivity is known to be size independent for 3-D materials, it remains an open question whether or not the thermal conductivity in 1D and 2D materials is size independent. As summarized in Sec. I, a divergent thermal conductivity scales with sample length $L$ (the positive power exponent $Z$), $k \sim L^z$, has been observed in most momentum conserving 1D toy lattice models, and also (up to a certain length) in some quasi 1D nanostructures. One the other hand, in the strict 2-D systems, as revealed by the Fermi-Pasta-Ulam (FPU) model, the thermal conductivity scales logarithmically with system size. Recent intensive first principles calculations for graphene (Fugallo et al. 2014), and graphene nanoribbons (Majjee and Aksamija 2016), indicate that the logarithmic relation seems to hold up to a certain length of ~1 mm, which is corroborated in experiments using the micro-bridge measurement (Xu*, et al.*, 2014a). These first principles calculations predict that, in order to observe a size independent thermal conductivity, one should utilize samples larger than 1mm,



which is a quite challenge task for experimentalists. In addition, because the crystal structure of graphene is quite different from most other 2-D crystals, it is not clear whether the thermal conductivity of these materials share the same graphene's length dependence.

Fig. 4(a, b) shows the experimental setup for thermal conductivity measurements of single-layer graphene with different sample length, from 300 nm to 10 μm by changing the separation between the heating and sensing membranes, where graphene is suspended on a micro-bridge device (Xu, *et al.*, 2014a). Fig. 4(c) summarizes the measured thermal conductivity of graphene as a function of sample size (Xu, *et al.*, 2014a), as well as from calculations including the non-equilibrium molecular dynamics simulations (NEMD) (Park, *et al.*, 2013; Mu, *et al.*, 2014; Xu, *et al.*, 2014a; Barbarino, *et al.*, 2015) and PBTE calculations (Fugallo, *et al.*, 2014; Lindsay, *et al.*, 2014; Gu and Yang, 2015; Majee and Aksamija, 2016). NEMD simulations have also been performed to mimic the micro-device based measurement by sandwiching the sample between a hot reservoir and a cold reservoir, where the boundary scattering is considered in these simulations. For PBTE calculations, a boundary scattering term that is proportional to the ratio between group velocity of phonon mode, $v_{\mathbf{q}s}$, and sample size, *L*, is included, so that phonons are more strongly scattered by the boundaries when the size of sample is small (Mingo and Broido, 2005). From all of these studies, it is evident that the thermal conductivity of graphene increases with sample size, for room temperature samples of up to tens of microns. The thermal conductivity of some other 2-D materials from PBTE also exhibits strong length dependence (Lindsay and Broido, 2011a; Gu and Yang, 2014; Zhu, *et al.*, 2014; Gu and Yang, 2015), as shown in Fig. 4(d).

To analyze the thermal conductivity of infinitely large 2-D crystals, ideally one needs to employ either the iterative approach (Omini and Sparavigna, 1996; Broido, *et al.*, 2005) or a



variational method (Fugallo, *et al.*, 2013; Fugallo, *et al*., 2014). The latter was developed most recently and is equivalent to the iterative approach but numerically more stable and obtains the full solution of the PBTE. The full solution has been shown to be crucial for predicting the phonon transport in 2-D materials, such as graphene, where the normal phonon scatterings are stronger than the Umklapp scatterings. Unlike the resistive Umklapp scatterings, the normal scattering processes do not contribute resistance to heat flow directly, but play important roles in redistributing the non-equilibrium phonon density. When normal scattering dominates, phonon transport deviates from the kinetic picture of a phonon gas obeying the relaxation time (mean free path) approximation, instead it exhibits hydrodynamic features due to the macroscopic drift motion of phonons (Cepellotti, *et al.*, 2015; Lee, *et al.*, 2015a). Such phenomenon can be correctly captured by the full solution of the PBTE. However, since the non-equilibrium phonon population of a specific phonon mode under a temperature gradient is affected by the non-equilibrium population of all other phonon modes in the crystal, the mode-specific thermal conductivity contributed by the phonon modes in the long-wavelength limit cannot be determined unless a (computationally infeasible) mesh with infinite points is used to sample the first Brillouin zone.

Instead of seeking full solution of PBTE, we examine the thermal conductivity using Eq. (1). This expression is based on the simple kinetic theory of phonons, or equivalently the solution of PBTE under single-mode relaxation time approximation (SMRTA), which uses phonon lifetimes as the phonon relaxation time. The phonon scattering rates for long-wavelength phonons can be determined analytically, as discussed below. Comparing to the full solution of PBTE, the SMRTA gives the lower bound value of the thermal conductivity (Fugallo, *et al.*, 2013). In other words, we would conclude that the thermal conductivity from a more accurate iterative solution



of PBTE is unbounded if the thermal conductivity calculation using Eq. (1) under SMRTA is found to be unbounded, but not vice versa. According to Eq. (1), the anomalous thermal conductivity comes from the long-wavelength phonons, or acoustic phonons near the Γ point, since the integrand in the equation at other $q$ vector is a finite number.

Here, we focus our discussion on hexagonal 2-D crystals, since a large portion of 2-D materials possess such a hexagonal crystal structure. Furthermore, the isotropic phonon dispersion of hexagonal crystals near the Γ point greatly simplifies this analysis. To obtain a finite thermal conductivity of a hexagonal 2-D crystal, the long-wavelength (small wave vector) phonons on each phonon branch should satisfy the condition $\omega_{\mathbf{q}s}^2 n_{\mathbf{q}s}^0 \left( n_{\mathbf{q}s}^0 + 1 \right) v_{\mathbf{q}s}^2 \tau_{\mathbf{q}s}^{\mathrm{ph}} \propto q^n$ with $n > -2$ (Gu and Yang, 2015).

While the in-plane acoustic phonons obey the relationship of $\omega_{\mathrm{TA,LA}} \propto q$ near the Γ point, flexural acoustic phonons depends strongly on the strain conditions in hexagonal 2-D materials. According to the elastic theory, the flexural acoustic phonon branch follows $\omega_{\mathrm{ZA}} \propto q^2$ when the sheet is free of stress (Zabel, 2001), which is regarded as a signature of the phonon dispersion curves of 2-D materials. However, some complicated theoretical studies predicted that the dispersion deviates from a simple quadratic dependence is due to mode renormalization (Mariani and von Oppen, 2008; Wang and Daw, 2016). If a 2-D material sheet is stretched, even with infinitesimal small strain, the dispersion is linearized, i.e., $\omega_{\mathrm{TA,LA}} \propto q$. Hence, a finite thermal conductivity requires that the relaxation time of long-wavelength acoustic phonon be $\tau_{\mathbf{q}s}^{\mathrm{ph}} \propto q^n$ with $n > -2$ if the dispersion is linear, or $n > -4$ when the dispersion is quadratic.

Recent progress in phonon transport modeling has made it possible to numerically determine the scattering rates for the three-phonon processes (Omini and Sparavigna, 1996; Broido, *et al.*,



2005; Esfarjani, *et al*., 2011), and thus the lifetime of each phonon mode. However, it is still cost-prohibitive to numerically distinguish how phonon lifetime depends on its wavevector as the wavelength approaches zero, since it requires a very fine mesh sampling of the Brillouin zone. Instead, if three-phonon scattering is organized into a series of scattering channels, one can derive analytical expressions for the relation between wave vector and scattering rate for each scattering channel (Bonini*, et al.*, 2012). This provides an efficient approach to find the dominant scattering channels for long-wavelength phonons and thus the dependence of phonon lifetime with its wavelength.

The scattering of a long-wavelength acoustic mode can be divided into four distinct processes: 1) the annihilation of one phonon mode with another one to generate a third phonon on the same branch, 2) the annihilation of one phonon mode with another one to generate a third phonon on a different branch; 3) the decay of one phonon to two lower-frequency phonon modes on the same branch, and 4) the decay of one phonon to two phonon modes on two lower acoustic branches. These processes are illustrated in <span style="color:red">Fig. 5</span>. Note that the scaling relation between the scattering rate and phonon wave vector can vary greatly with strain and crystal structure.

Take unstrained graphene as an example, Bonini *et al*. (Bonini*, et al.*, 2012) analytically derived the scattering rates, or the inverse of the phonon lifetimes, for both the in-plane acoustic phonons and the flexural acoustic (ZA) phonons. They found that the long-wavelength in-plane acoustic phonons are predominantly scattered through a decay process that splits one in-plane acoustic phonon into two flexural acoustic phonons due to the quadratic shape of the ZA branch, and the corresponding scattering rate is constant. ZA phonons also scatter with in-plane acoustic phonons in the reverse manner: they are annihilated along with an in-plane acoustic phonon to generate a new in-plane acoustic phonon. The scattering rate of ZA modes due to such events



scales as q², when *q* approaches zero (See Fig. 6(a)). Therefore, no matter how phonons are scattered through other scattering channels, the scattering rates of both in-plane acoustic phonons and flexural phonons would satisfy the conditions that ensures a finite thermal conductivity. The equilibrium molecular dynamics (EMD) simulations of Pereira and Donadio (Pereira and Donadio, 2013) also showed that stress-free graphene has a finite thermal conductivity. They found that the thermal conductivity is converged as the size of simulation domain increases. However, Lindsay (Lindsay, 2010) pointed out that the thermal conductivity becomes unbounded when using the iterative solution of PBTE. The underlying mechanisms are still not well understood. Since the main scattering mechanism for long-wavelength ZA phonon is the normal process that two ZA phonons annihilate to generate an in-plane acoustic phonon, ZA+ZA<->LA/TA, (Bonini*, et al.*, 2012), the unbounded thermal transport could be related to the fact that a large portion of the in-plane phonons is converted to ZA phonons.

When a tensile strain, even an infinitesimal one, is applied to a single layer graphene, the dispersion of the ZA phonon branch becomes linearized (Bonini*, et al.*, 2012). In this case, the ZA phonons cannot be as efficiently scattered through the processes where two ZA modes are annihilated into a TA or LA mode (a common process in unstrained graphene), and the scattering rate of long-wavelength ZA phonons scales as $q^3$ (Bonini*, et al.*, 2012). Unlike in unstrained graphene, annihilation processes involving ZO phonons become dominant in the long-wavelength limit for strained graphene, whose inverse of scattering rates, or phonon lifetimes, are proportional to $q^{-2}$ (See Fig. 6(b)). Thus, in the long-wavelength limit, the condition $\tau_{\mathbf{q}s}^{\mathrm{ph}} \propto q^n$ with *n* > -2, does not hold for ZA phonons, and the thermal conductivity of a strained sample diverges logarithmically as the length of the sample increases.



For 2-D materials with more than one-atom layer, such as silicene, the symmetry selection rules for the scattering of flexural phonons no longer apply (Gu and Yang, 2015). The scattering of ZA phonons becomes much more frequent than that in the one-atom-thick 2-D materials. Numerical calculations showed that the thermal conductivity of infinitely large unstrained silicene, germanene and stanene sheets has finite values from both a SMRT solution and iterative solution of PBTE (Kuang, et al., 2016).

If the dispersion of ZA branch is *linear,* the scattering rate of ZA phonons is found to scale as $q^{-1}$ near the $\Gamma$ point. As a result, the unbounded thermal conductivity does not come from the ZA phonons (as in strained one-atom-thick 2-D materials), instead it comes from the long-wavelength LA phonons. According to the analytical derivation of Gu and Yang (Gu and Yang, 2015), only the annihilation process of a long-wavelength acoustic phonons with two neighboring modes on the same branch could make the inverse of scattering rate follow $q^n$ with $n > -2$. For a long-wavelength LA phonon $\mathbf{q}s$, its frequency can be expressed as $\omega_{\mathbf{q}s} = v_{LA}q$, with the sound velocity of LA branch $v_{LA}$. The frequencies of the two phonons, $\mathbf{q}'s'$ and $\left(\mathbf{q} + \mathbf{q}'\right)s''$ on the same branch that can scatter with $\mathbf{q}s$ have to satisfy the condition, $\omega_{(\mathbf{q}'+\mathbf{q})s'} - \omega_{\mathbf{q}'s'} = v_{LA}q$. As $q$ approaches to zero, the condition becomes $\mathbf{v}_{\mathbf{q}'s'} \cdot \mathbf{q} = v_{LA}q$. However, since the sound velocity of LA branch is usually larger than the group velocity of other phonon modes, long-wavelength LA phonons cannot be scattered by two phonons on the same branch. Thus, the thermal conductivity of silicene with a linear ZA branch diverges, which is not because of the ZA modes but because of the in-plane LA modes. The divergent thermal conductivity of silicene with respect to the sample size was also confirmed by other calculations (Kuang, *et al.*, 2016; Xie, *et al.*, 2016). It is worth mentioning that in the first-principles calculations, due to numerical



inaccuracy, the shape of flexural phonons of unstrained sheet near the $\Gamma$ point could be linear, which leads to an unbounded thermal conductivity for infinitely large sheet. By imposing additional invariances on the extracted second-order harmonic force constants from first-principles, the quadratic dispersion of flexural phonons can be recovered (Carrete, *et al.*, 2016).

It should be noted that the above discussion considers only three-phonon scattering processes in hexagonal lattices. Although the higher-order phonon scatterings beyond three-phonon scatterings are usually considered to be weaker than the three-phonon scatterings (Ecsedy and Klemens, 1977), they might play an important role in scattering long-wavelength phonons due to the very large scattering phase space. Given the ultra-low bending stiffness of atomic-thin 2-D materials (Wei *et al.*, 2012a), ripples can easily be formed in those materials under thermal undulation (Fasolino et al. 2007). The ripples could serve as additional channels to scatter phonons. For example, the intrinsic ripples that occur in suspended graphene makes the symmetry selection rule break down, leading to stronger phonon scatterings and thus reduces the thermal conductivity. From a rough estimation, one might expect that the corresponding strength enhancement of three-phonon processes is weak (Lindsay, et al., 2010b), but the additional scattering might alter the scaling relation between phonon wave vector and scattering rate for long-wavelength phonons, thus changing the length dependence behavior of the thermal conductivity. While 2-D hexagonal crystals have an isotropic dispersion near the $\Gamma$ point, this is not the case for many other 2-D crystal structures. Therefore, the scattering phase spaces of long-wavelength acoustic phonons in non-hexagonal crystals can be quite different from those in hexagonal crystals. For instance, a large portion of long-wavelength LA phonons could scatter with two modes that are on the same branch, as the group velocities of these LA modes are not



the largest one among all phonon modes, which is different from hexagonal crystals with a linear ZA branch.

## C. Geometrical effects (nanoribbons and few-layers)

Width and thickness play an important role not only in determining electronic properties of layered 2-D materials, they can also influence the phonon scattering rate thus altering a material's thermal properties. In conventional materials, when a natural length scale (width, length, or thickness) becomes smaller than the phonon mean free path, the increased phonon-boundary scattering will reduce thermal conductivity. However, when the sample size becomes comparable to the phonon wavelength, phonons are confined to 1-D ribbons or 2-D planes and the effective thermal conductivity might not decrease monotonically with reduced size. Size-dependent thermal conductivity in 2-D crystals could be quite different from that of conventional thin films. In this section, we review recent studies on thermal transport in nanoribbons and few-layer 2-D materials, and discuss the possible origins of the size dependence.

### C.1 Width dependence

The phonon properties and thermal conductivity of nanoribbons of 2-D materials (especially graphene nanoribbons) has been extensively studied, primarily through molecular dynamics (MD) simulations (Guo, *et al.*, 2009; Hu, *et al.*, 2009; Evans, *et al.*, 2010; Ye, *et al.*, 2015, Majee and Aksamija 2016), PBTE-based simulations (Aksamija and Knezevic, 2011; Nika, *et al.*, 2009a; Nika, *et al.*, 2009b; Nika, *et al.*, 2011) and atomistic Green's function theory calculations (Xu, *et al.*, 2009; Tan, *et al.*, 2010). From most MD and PBTE calculations, the thermal conductivity of graphene nanoribbons is found to be lower than that of larger graphene sheets and it decreases



with decreasing ribbon width, although the degree of reduction depends upon the theoretical methods and interatomic potentials employed. The lower thermal conductivity of graphene nanoribbons compared with larger graphene sheets was later confirmed by a few experimental measurements. For instance, thermal conductivity of $SiO_2$-supported graphene nanoribbons with different lengths and widths was measured using the micro-bridge method (Bae, *et al.*, 2013). The thermal conductivity of single-layer graphene nanoribbons with a length $L = 260$ nm at room temperature was found to increase rapidly with the increase of the nanoribbon width when the width is below 100 nm. Measurements of suspended graphene nanoribbons with sub-micron length and width were performed using the electrical self-heating method (Xie, *et al.*, 2013; Li, *et al.*, 2015), where the thermal conductivity of graphene nanoribbons was found to be only several hundred W/(m K) at room temperature, significantly lower than that of the large suspended graphene sheet.

Despite most work showing that narrow nanoribbons have lower thermal conductivity than wider ones, the atomistic Green's function calculations show that the normalized thermal conductance (the ratio between thermal conductance and the cross-sectional area of a nanoribbon) decreases with width for many nanoribbons of 2-D materials, such as graphene (Xu, *et al.*, 2009; Tan, *et al.*, 2010), boron nitride (Ouyang, *et al.*, 2010) and graphyne (Ouyang, *et al.*, 2012). This is due to the harmonic approximation assumed in the atomistic Green's function theory calculations. Atomistic Green's function calculations only provide an upper (ballistic) limit for the thermal conductivity and do not necessarily ensure the correct width dependence for the thermal conductivity of nanoribbons unless the length of the nanoribbon is much smaller than the phonon mean free path. Indeed, the sample length can also play a role in determining the width-dependent thermal conductivity of nanoribbons (Nika, *et al.*, 2012).



Many explanations have been explored to interpret the width-dependent thermal conductivity of graphene nanoribbons. In general, there are three possible explanations: phonon-edge (boundary) scattering, phonon confinement (Tan*, et al.*, 2010) and phonon localization (Wang*, et al.*, 2012b). In the first explanation, the phonon modes of nanoribbons are assumed to be the bulk phonon modes of a large 2-D sheet. These bulk phonons are occasionally scattered by the edges (in addition to the intrinsic phonon-phonon scattering), leading to a lower thermal conductivity than that of a large 2-D sheet. The effect becomes more pronounced for a narrower ribbon with rougher edges. Such an explanation has been widely used to explain the lower thermal conductivity of nanostructured bulk semiconductors (Yang and Chen, 2004; Yang*, et al.*, 2005). In the second explanation, the nanoribbon is treated as a quasi-1-D crystal, which has quite different phonon dispersions and scattering rates from those of a large graphene sheet. In particular, the phonon dispersion and group velocities of nanoribbons with different widths differ substantially, resulting in width-dependent thermal transport. In the third explanation, the thermal conductivity reduction of nanoribbons is attributed to phonon localization. The vibration of atoms in the edge is likely to be localized and decoupled from other atoms within the nanoribbon. Localized phonons do not participate in energy transfer.

In addition to the width of nanoribbons, several other factors affecting the thermal conductivity have been identified through numerical simulations, including edge chirality, roughness, and hydrogen-passivation. Both MD simulations and Green's function calculations showed that a zigzag nanoribbon has a larger thermal conductivity than an armchair nanoribbon of an equal width. According to the phonon-edge scattering interpretation, boundary scattering is expected to be weaker in zigzag nanoribbons because the edge roughness of zigzag nanoribbons is less than that of armchair ones, i.e. the armchair nanoribbon has more edge scatterers per unit



length (Haskins, *et al.*, 2011). This hypothesis was confirmed by wave-packet MD, where the propagation of phonons is monitored before and after scattering with the edges (Wei, *et al.*, 2012b). There were also explanations from the viewpoint of phonon confinement (Xu, *et al.*, 2009; Tan, *et al.*, 2010; Ye, *et al.*, 2015) and localization (Wang, *et al.*, 2012b).

Aside from the intrinsic zigzag and armchair edges, the thermal conductivity of nanoribbons could be further reduced by introducing larger scale edge roughness (Evans, *et al.*, 2010; Aksamija and Knezevic, 2011; Haskins, *et al.*, 2011) or hydrogen-passivation (Evans, *et al.*, 2010; Hu, *et al.*, 2010), which provide additional mechanisms for tuning the thermal conductivity of 2-D materials. For example, compared to zigzag nanoribbons of the same width, an introduction of a root mean squared (RMS) roughness of 0.73 nm, can reduce the thermal conductivity of a 15-nm-wide graphene nanoribbon by 80%. Hydrogen passivation reduces the thermal conductivity by around 50% and 25% for zigzag and armchair graphene nanoribbons with a width of 1.5 nm (Hu, *et al.*, 2010), respectively.

While the thermal conductivity of many 2-D material at around room temperature roughly follows a $1/T$ temperature dependence (which is regarded as a sign of strong Umklapp phonon-phonon scattering), the thermal conductivity of graphene nanoribbons shows a different trend, where thermal conductivity linearly increases or remains almost unchanged with temperature (Hu, *et al.*, 2009; Xie, *et al.*, 2013). Such temperature dependence should be attributed to the strong phonon-boundary scattering, similar to that in the conventional 3-D materials, whose scattering rate is insensitive to temperature. It was very recently proposed that at cryogenic temperature (and even up to room temperature) the anomalous thermal conductivity might be due to second sound (Cepellotti, *et al.*, 2015; Lee, *et al.*, 2015a), where the effective temperature along the width direction is not uniform but quadratic, just like the Poiseuille flow velocity



profile of fluids in tubes or channels, which is attributed to the fact that normal phonon-phonon scattering processes dominate over resistive Umklapp phonon-phonon scattering.

The thermal conductivity of nanoribbons of a few other 2-D materials has also been reported, including boron nitride (Sevik, *et al.*, 2011) and MoS$_2$ (Jiang, *et al.*, 2013; Liu, *et al.*, 2013; Liu, *et al.*, 2014c) using MD simulations. Liu et al. (Liu, *et al.*, 2013) found that the thermal conductivity is almost independent of nanoribbon width. This might be due to the strong anharmonicity in the empirical potentials employed in their MD simulations. It should always be kept in mind that the calculated thermal conductivity depends strongly on the empirical potentials employed when interpreting the MD results.

Due to the low thermal conductivity of nanoribbons, thermoelectric performance of nanoribbons of many materials has also been studied. Different strategies have been proposed to further suppress the thermal conductivity while maintaining good electronic transport (Nguyen, et al., 2014; Hossain, et al., 2014; Tran, et al., 2017), for example, introducing defects in nanoribbons and constructing junctions of two different 2D materials.

## C.2 Layer thickness dependence

The thermal conductivity of thin films of a few layered 2-D materials, such as black phosphorus (Jang, *et al.*, 2015; Lee, *et al.*, 2015b; Luo, *et al.*, 2015; Zhu, *et al.*, 2016b), TaSe$_2$ (Yan, *et al.*, 2013) and Bi$_2$Te$_3$ (Pettes, *et al.*, 2013), with a thickness in the order of 10 to 100 nm, has been measured. The thermal conductivity of these layered 2-D materials was reported either to be substantially lower than that of their bulk counterparts or to increase with sample thickness. This observation seems to support the argument that the classic size effects in conventional 3-D materials, which assumes that boundary scattering reduces the thermal conductivity, also occur



in layered 2-D crystals. However, when the thickness is further reduced to around 10 nm, the dependence of thermal conductivity on sample thickness becomes quite complicated, likely due to the change of phonon dispersion. While Raman spectra of a variety of few-layer 2-D materials have provided the evidence of thickness-dependent phonon properties (Ferrari, *et al.*, 2006;  Li, *et al.*, 2012a;  Castellanos-Gomez, *et al.*, 2014), the thickness dependence of thermal conductivity in these materials is not conclusive yet due to the possible change of both phonon dispersion and phonon scattering. Here, we present a brief summary of these studies on thickness dependent thermal conductivity of a few layered 2-D materials, including graphene, $MoS_2$, $Bi_2Te_3$ and black phosphorus.

Fig. 7(a) shows the thermal conductivity of graphite, few-layer graphene and single layer graphene. The thermal conductivity of graphite is measured to be about 2000 W/(m K) at room temperature, while that of single layer graphene ranges from 2500 to 5000 W/(m K). The theoretical studies using both NEMD (Wei, *et al.*, 2011b) and PBTE (Lindsay, *et al.*, 2011; Singh, *et al.*, 2011;  Kuang, *et al.*, 2015) confirmed that single-layer graphene has a much higher thermal conductivity than that of graphite. As discussed in Sec. II. A, this is because phonon scattering in single-layer graphene has to obey the symmetry selection rule (Lindsay, *et al.*, 2010b), which leads to a smaller scattering phase space and weaker scattering for the flexural (ZA) phonon modes. Some experiments (Ghosh, *et al.*, 2010) and most of the numerical simulations (Lindsay, *et al.*, 2011;  Singh, *et al.*, 2011;  Kuang, *et al.*, 2015) further showed that thermal conductivity gradually decreases when the number of layers is increased, while some other experiments reported much smaller thermal conductivity (Pettes, *et al.*, 2011) due to polymer residues on the samples. In encased graphene, the inverse thickness dependence of the thermal conductivity has been observed. (Jang, *et al.*, 2013).



Theoretical study showed that additional ZA-like low-frequency optical phonon branches are generated in multi-layer graphene. Compared to the ZA modes in single-layer graphene, these ZA-like phonons in multilayer graphene have lower average phonon group velocities and much larger phonon scattering rates due to the greater number of scattering channels available, which results in a lower thermal conductivity. A similar monotonic reduction of thermal conductivity with respect to the number of layers was also reported for graphene nanoribbons from MD simulations (Zhong, *et al.*, 2011; Cao, *et al.*, 2012) and boron nitride from PBTE-based calculations (Lindsay and Broido, 2012), which could be explained by the similar reasons revealed for graphene.

Compared to graphene, recent experiments showed a quite different trend for the layer thickness-dependent thermal conductivity of $MoS_2$, as summarized in Fig. 7(b) and Table I. While the measured basal-plane thermal conductivity of single-layer $MoS_2$ by different researchers differ – due to differences in sample quality and experimental conditions – the thermal conductivity of $MoS_2$ with more than 4 layers seems to follow an opposite trend from that of graphene, i.e. the thermal conductivity of $MoS_2$ increases with the number of layers. A recent first-principles-based PBTE study (Gu, *et al.*, 2016) showed that the basal-plane thermal conductivity of 10-μm-long samples reduces monotonically from 138 W/(m K) to 98 W/(m K) for naturally occurring $MoS_2$ when its thickness increases from one layer to three layers, and the thermal conductivity of tri-layer $MoS_2$ approaches to that of bulk $MoS_2$. The reduction is attributed to both the change of phonon dispersion and the thickness-induced anharmonicity. Phonon scattering for ZA modes in bi-layer $MoS_2$ is found to be substantially larger than that of single-layer $MoS_2$, which is attributed to the fact that the additional layer breaks the mirror symmetry as discussed in Sec. II.A. In Fig. 7(b), the measured thermal conductivity of $MoS_2$ is



also presented (Sahoo, *et al.*, 2013; Jo, *et al.*, 2014; Liu, *et al.*, 2014b; Yan, *et al.*, 2014; Zhang, *et al.*, 2015b), but the experimental results show no clear thickness dependence.

Bi$_2$Te$_3$ is another interesting layered 2-D material, which has been widely studied in the fields of thermoelectrics (Venkatasubramanian, *et al.*, 2001; Poudel, *et al.*, 2008) and topological insulators (Chen, *et al.*, 2009; Zhang, *et al.*, 2009). As shown in Fig. 7(c), EMD study (Qiu and Ruan, 2010) on few-layer Bi$_2$Te$_3$ showed that the thermal conductivity of Bi$_2$Te$_3$ does not change monotonically with thickness, but exhibits a minimum at a thickness of three quintuples. At a thickness of 5 nm (five quintuples), the thermal conductivity recovers to that of the bulk counterpart. Such a non-monotonic layer thickness dependence was attributed to the competition between phonon-boundary scattering and the Umklapp phonon-phonon scattering in the thickness range they explored (Qiu and Ruan, 2010). However, such a theoretical prediction seems to be conflict with the micro-bridge measurement for Bi$_2$Te$_3$ thin films of 9 nm to 25 nm thickness (Pettes, *et al.*, 2013). Pettes showed that the thermal conductivity roughly increases with thickness. Note that the experimental measurements by Pettes are on much thicker samples, while the calculation by Qiu and Ruan (Qiu and Ruan, 2010) are for thin samples with only a few layers.

Fig. 7(d) shows the measured thermal conductivity of black phosphorus along both the zigzag and the armchair directions from different research groups (Jang, *et al.*, 2015; Lee, *et al.*, 2015b; Luo, *et al.*, 2015; Zhu, *et al.*, 2016b; Sun, *et al.*, 2017). Interestingly, in both directions the measured thermal conductivity of black phosphorus thin films increases with the thickness, although the data from different groups vary noticeably. While the PBTE prediction for bulk black phosphorus agrees well with time-domain thermoreflectance (TDTR) measurements, there is still a large discrepancy among the predicted thermal conductivities of single-layer black



phosphorene (Zhu, *et al.*, 2014;  Jain and McGaughey, 2015;  Qin, *et al.*, 2016b), due to the choice of the cutoff of interatomic forces and whether the van der Waals interaction is considered in the first-principles calculations (Qin, *et al.*, 2016b).

From the above summary on the thickness-dependent thermal conductivity of several layered 2-D materials, it can clearly be seen that there is a huge gap between the theoretically calculated results and the experimental measurements. The calculated thermal conductivity of 2-D materials with one or few layers is usually higher than that of the bulk counterpart, while the measurement usually does not follow such a trend. Polymer residues, absorbed gas, and oxidization are considered to contribute to this discrepancy. To better understand the difference between the theoretical calculations and measured data, more efforts on numerical simulations should be given to the study of thermal properties of more realistic materials in order to see how imperfections, which are common in experiments, affect the thermal conductivity of 2-D materials.

Apart from stacking 2-D layers based on their bulk forms, for example, AB stacking for graphene and $MoS_2$, it is possible to stack different layers in other ways to achieve few-layer pristine 2-D materials with novel thermal properties. Nika et al. and Cocemasov et al. (Nika, *et al.*, 2014;  Cocemasov, *et al.*, 2015) theoretically studied the phonon properties and the in-plane thermal conductivity of twisted bi-layer graphene. The Raman-based optothermal measurement confirmed that the thermal conductivity of such bi-layer graphene could be substantially lower than that of the bi-layer graphene with AB stacking (Li, *et al.*, 2014). Stacking the exfoliated $Bi_2Te_3$ thin films to form the "pseudosuperlattice" has been proven as an effective method to reduce the thermal conductivity from the bulk value (Goyal, *et al.*, 2010;  Teweldebrhan, *et al.*,



2010). The thermal conductivity reduction is thought to be due to the rough boundary scattering or spatial confinement effect of acoustic phonons (Teweldebrhan, *et al.*, 2010).

## D. Strain effects

Strain and deformation commonly exist in 2-D materials. When 2-D materials are integrated into nano-devices, typically they are deformed due to constraints imposed during device assembly. When two or more kinds of 2-D crystals are vertically stacked or laterally connected to form heterostructures, 2-D materials are likely to be strained due to the lattice mismatch. On the other hand, strain engineering has also been proposed to tune electronic, photonic and even thermal properties. (Mohiudin, et al. 2009, Conley, et al. 2013, Fan, et al. 2017).

The studies on strain-dependent thermal conductivity of solids and liquids can be traced back to 1970s and 1980s (Ross, *et al.*, 1984), where strain was usually induced by pressure (compressive stress). The experimental demonstration of utilizing strain to actively control phonon and thermal properties has been applied to many different kinds of material systems (Hsieh, *et al.*, 2009; Hsieh, *et al.*, 2011; Alam, *et al.*, 2015). Theoretical studies have also been performed to understand the origins of strain-dependent thermal conductivity of both bulk crystals (Picu, *et al.*, 2003; Bhowmick and Shenoy, 2006; Li, *et al.*, 2010; Parrish, *et al.*, 2014) and nanostructured materials (Xu and Li, 2009; Xu and Buehler, 2009; Li, *et al.*, 2010). While it is necessary to quantify the strain dependence of thermal conductivity, it is still more challenging to precisely control the exerted strain in 2-D materials than in 3-D materials. Therefore, the study of strain-dependent thermal conductivity of many 2-D materials has been limited to just numerical simulations.



Li *et al.* conducted EMD simulations (Li, *et al.*, 2010) to show that the thermal conductivity of graphene monotonically decreases with the increase of tensile strain, similar to most bulk materials. However, the calculations of Fan *et al.* (Fan, et al. 2017) using both EMD and NEMD showed an increase in thermal conductivity with the increase of tensile strain, particularly due to the contributions from flexural modes. Recent calculations using PBTE-based approaches give quite different trends for strain dependence. For example, under RTA, Bonini, *et al.* (Bonini, *et al.*, 2012) found that the thermal conductivity of strained graphene is unbounded, due to the weak scattering for ZA phonons, and the tensile strain enhances the thermal conductivity in the strain range they explored for an infinite large graphene sheet. Some other researchers (Fugallo, *et al.*, 2014) showed that the thermal conductivity of graphene is a finite value when the PBTE is solved iteratively. The tensile strain is also found to either enhance or reduce the thermal conductivity of graphene depending on the sample size (Fugallo, *et al.*, 2014; Kuang, *et al.*, 2016a). Many of these discrepancies indeed come from computational details. In MD simulations, due to the finite simulation domain size, the phonons whose wavelengths are larger than twice of the simulation domain size do not exist in the simulation system, and thus do not contribute to the thermal conductivity. As the long-wavelength phonons play unique roles in 2-D crystals (as discussed in Sec. II. B), the thermal conductivity results from MD simulations could be sensitive to both the size of the simulation domain and the treatments employed to compensate for the thermal conductivity from the long-wavelength phonons. For PBTE calculations, since a finite number of $\mathbf{q}$-points are used for numerical integration over the Brillouin zone where the heat conducted by long-wavelength phonon modes near the first Brillouin zone is not counted and the calculated thermal conductivity exhibits a strong q-mesh dependence, just like the simulation domain size effects in MD simulations. The strain



dependence on thermal conductivity from PBTE calculations might not be justified if the calculations are performed using a **q**-mesh with fixed number of **q**-points. Special attention needs to be paid to the computational details when interpreting these simulation results.

Fig. 8(a) shows that the strain dependence of thermal conductivity in other single-layer 2-D materials is quite diverse. A monotonically reduction of thermal conductivity was observed for single-layer $MoS_2$ from MD simulations (Jiang, *et al.*, 2013) and for penta-graphene from PBTE (Liu, *et al.*, 2016). The thermal conductivity of some buckled 2-D crystals, e.g. silicene (Kuang, *et al.*, 2016b; Xie, *et al.*, 2016) and penta-$SiC_2$ (Liu, *et al.*, 2016), and few-layer 2-D materials, such as graphene and boron nitride (Kuang, *et al.*, 2015), exhibit a non-monotonic (up-and-down) relation with strain. Penta-$SiN_2$ was discovered from a first-principles-based PBTE calculations to have a very unusual thermal conductivity dependence on strain, which is increased by an order of magnitude at a tensile strain of 9% (Liu, *et al.*, 2016).

The diverse dependence of thermal conductivity on strain for different 2-D materials can be attributed to the different strain dependence of heat capacity, group velocity, and phonon-phonon scattering rates for 2-D sheets with different crystal structures in accordance with the simple kinetic theory. In general, the stretching of a material could lead to a downshift of the phonon spectra for many 3-D materials. This is also true for most 2-D materials. Fig. 8(b) shows phonon dispersion of silicene as an example (Xie, *et al.*, 2016). In most of the calculated phonon dispersions of tensile stained 2-D, we observe optical phonon softening (Liu, *et al.*, 2007; Li, 2012; Xie, *et al.*, 2016). For acoustic phonons, the frequency of the longitudinal and transverse acoustic phonons shows a similar reduction, but the flexural acoustic phonons behave differently near the $\Gamma$ point due to phonon stiffening. Therefore, tensile strain could either increase or



reduce the phonon group velocities. When the strain is large, the downshift of phonon spectra might be significant and the velocity of most phonon modes is thus reduced.

Apart from the phonon dispersion, lattice distortion (including phase change) could be another reason for the diverse responses of thermal conductivity when a tensile strain is applied. For many non-one-layer-thick 2-D crystals, such as those with buckled structure, tensile strain could substantially reduce the buckled distance, e.g. silicene (Xie, *et al.*, 2016), or even eliminate the buckling, e.g. penta-$SiN_2$ (Liu, *et al.*, 2016). As discussed in Sec. II.B, the buckled structure could induce strong scattering for ZA phonons (as well as LA and TA phonons). Thus, applying tensile strain weakens phonon scattering by partially recovering the selection rule of symmetry. The interplay between the shift of phonon spectra and the change of phonon scattering channels results in the non-monotonic strain dependence of thermal conductivity for many 2-D materials.

For example, the up-and-down strain-dependent thermal conductivity for silicene was observed from PBTE calculations (Kuang, *et al.*, 2016b; Xie, *et al.*, 2016) (See Fig. 8(a)). The thermal conductivity increased first to about 8 times the unstrained state's at 4-8% tensile strain, then it decreased to around 10-50% the maximum thermal conductivity when the tensile strain was further increased to 10%. When tensile strain is applied, the strain-induced flattening of the structure weakens many of the dominant scattering channels in unstrained silicene, such as ZA+ZA->ZA, ZA-> ZA+ZA, TA+ZA->LA/TA, resulting in an increased thermal conductivity. These suppressed channels correspond to totally forbidden channels in the purely planar 2-D crystals (e.g. graphene). The sudden jump of thermal conductivity of 2-D penta-$SiN_2$ at an 8% strain, shown in Fig. 8(a), is due to the transition of penta-$SiN_2$ from a buckled lattice to a flat one.



Compressive strain also greatly affects the thermal conductivity of 2-D crystals. According to EMD simulations (Li, *et al.*, 2010), an 8% compressive strain reduces the thermal conductivity of graphene by 50% relative to its unstrained state. The decrease of thermal conductivity with respect to compressive strain is attributed to compression induced rippling of graphene, since rippling could induce additional phonon scattering channels in addition to the intrinsic phonon-phonon scattering. Similar compressive strain dependence of thermal conductivity is observed in other MD simulations (Guo, *et al.*, 2009; Wei, *et al.*, 2011a). Lindsay and Broido (Lindsay, *et al.*, 2010a) studied the thermal conductivity of carbon nanotubes with different diameters using the Boltzmann transport equation and showed that the thermal conductivity of graphene is larger than that of carbon nanotubes with different diameters due to the breakdown of symmetry selection rule in CNTs. Although phonon transport in carbon nanotubes is not exactly the same as rippled graphene, this theoretical study explicitly shows that the thermal conductivity is highly affected by the curvature of 2-D sheets. Compared with graphene, studies on other 2-D materials are relatively scarce, except very limited few based on MD simulations, e.g. Ding, *et al.*, 2015 (Ding, *et al.*, 2015).

## E. Device geometry (effect of substrates)

When 2-D materials are integrated for device applications, the heat dissipation channels and thermal properties of 2-D materials change dramatically from the suspended ones due to their contact with a substrate. While the heat generated in a suspended 2-D material has to flow within the suspended sample and then dissipate to the substrate at the contact regions, in a device geometry heat could simultaneously flow in the sample and dissipate to the substrate directly by penetrating across the interfaces between 2-D material and the substrate. Therefore, for a good



understanding of heat dissipation in devices it is crucial to know both the thermal conductivity of supported 2-D material and the interfacial thermal conductance between 2-D materials and substrates. In this section, we first discuss the change of thermal conductivity of 2-D materials when they are supported, followed by the interfacial thermal conductance between 2-D material and substrates.

### E.1 Thermal conductivity of supported 2-D materials

The basal-plane thermal conductivity of a supported 2-D crystal is very different from a suspended one, even though in most cases the interaction between the 2-D materials and the substrates is weak (typically is of van der Waals type). Pioneering work on the effect of substrates on phonon transport in 2-D materials was done by Seol *et al.* (Seol, et al., 2010), where the thermal conductivity of supported graphene on silicon dioxide substrate was measured using the micro-bridge method. The measured thermal conductivity is only around 600 W/(m K) as shown in Fig. 9(a). This is very close to the value predicted by NEMD simulations (Chen, *et al.*, 2013), where monolayer graphene is placed on top of a $SiO_2$ substrate. The thermal conductivity of graphene on a $SiN_x$ substrate has been systematically studied by Wang (Wang, 2013); and was found to be only around 430~450 W/(m K). As the number of 2-D material layers increases, the thermal conductivity increases as shown in Fig. 9(b). Other experimental measurements, including those using optothermal Raman method (Cai, *et al.*, 2010), 3-ω heat spreader (Jang, *et al.*, 2013) and frequency domain thermoreflectance (FDTR) measurement (Yang, *et al.*, 2014), also confirmed thermal conductivity reduction of graphene when it is placed on top of a silicon oxide substrate or sandwiched between two substrates. Taube *et al.* (Taube, *et*



*al.*, 2015) measured the supported 2-D MoS$_2$ using optothermal Raman method, and showed a lower thermal conductivity value when compared with that of suspended samples.

Substrates affect the thermal conductivity of supported samples in two distinct ways: changing the phonon dispersion and enhancing the phonon scattering rate. Raman spectra measurements provide direct evidence for substrate effects on optical phonons (Berciaud*, et al.*, 2008). Many theoretical studies were devoted to deepening the understanding of phonon transport in supported 2-D materials. The phonon-substrate scattering rates have been derived using the scattering matrix method, where the effects of substrate are modeled as point defects to the 2-D crystal (Seol*, et al.*, 2010). Under a harmonic approximation and assuming that the phonon dispersion of the supported graphene is identical to the freestanding one, Seol*, et al.* found that phonon scattering increases with the average van der Waals interatomic force constant between the graphene and the SiO$_2$ support increases. This model qualitatively explains the thermal conductivity reduction of supported graphene. A few MD simulations were performed to predict the thermal conductivity of supported 2-D materials (Ong and Pop, 2011; Chen and Kumar, 2012; Qiu and Ruan, 2012; Chen*, et al.*, 2013; Wei*, et al.*, 2014a). Spectral energy density analysis shows that the relaxation time of acoustic phonons is reduced more pronouncedly than that of optical phonons when graphene is placed on top of a silicon oxide substrate, and the flexural acoustic (ZA) phonons are most affected (Qiu and Ruan, 2012). Due to the renormalization of ZA phonons, the frequency of ZA phonon is no longer zero at the $\Gamma$ point, but is a finite value and the scattering of ZA phonons is enhanced (Ong and Pop, 2011). The shift of phonon dispersion also contributes to the interesting length dependence of the thermal conductivity in graphene. As predicted by NEMD simulations (Chen*, et al.*, 2013), the thermal conductivity of supported graphene is almost unchanged when the length is larger than 100 nm



whereas in suspended graphene it increases even when the length is larger than 1 micron. Chen et al (Chen, *et al.*, 2016) studied thermal conductivity of supported black phosphorus using NEMD simulations and found that the degree of anisotropy becomes larger when supported.

Intuitively, a stronger coupling of 2-D crystals to the substrate would lead to a larger reduction in thermal conductivity, since the vibration of 2-D materials would be more severely affected, as indicated by the phonon-substrate scattering rates derived using the scattering matrix method (Seol, *et al.*, 2010). Most MD simulations confirmed such a relationship between the coupling strength and the thermal conductivity for supported graphene (Chen and Kumar, 2012; Qiu and Ruan, 2012; Chen, *et al.*, 2013; Wei, *et al.*, 2014a). Some MD simulations, however, showed that this is not necessarily true. For example, Ong and Pop (Ong and Pop, 2011) found that when the strength of graphene-substrate interaction is increased by two orders of magnitude, heat flow along supported graphene, (or equivalently its thermal conductivity) increases by up to three times. They found that the strong coupling between quadratic ZA modes of graphene and the surface waves of the substrate leads to a hybridized linear dispersion with a reduced scattering rate and higher phonon group velocity.

The thermal conductivity of supported 2D materials depends not only on the substrate coupling strength described above, but also on the nature of substrates. A recent study by Zhang et al. (Zhang, *et al.*, 2017) found that a substrate like h-BN (which has a very similar structure to graphene) does not significantly reduce the thermal conductivity of graphene.

Since the thermal conductivity of most 2-D materials is suppressed when they are weakly bonded to a substrate, it is desirable to identify the thickness (layer number) beyond which the thermal conductivity of layered 2-D material thin films recover to the free-standing sample's higher value. Fig 10 shows the layer thickness-dependent thermal conductivity of both supported



and encased few-layer graphene. In both measurements (Jang, *et al.*, 2013; Sadeghi, *et al.*, 2013; Yang, *et al.*, 2014) and NEMD simulations (Chen, *et al.*, 2013), the thermal conductivity is found to increase with the number of atomic layers, in contrast to the suspended graphene where the thermal conductivity decreases with the layer number. Encased 10-layer graphene has a thermal conductivity larger than 1000 W/(m K) (Jang, *et al.*, 2013), and the supported graphene of 34 layers is found to fully recover to the bulk value of graphite (Sadeghi, *et al.*, 2013). The large thickness required to ensure a higher thermal conductivity of encased and supported graphene is probably due to the long mean free paths of phonons in graphite along the cross-plane direction, which was revealed recently by MD simulations (Wei, *et al.*, 2014b) and confirmed by experimental measurements (Fu, *et al.*, 2015; Zhang, *et al.*, 2016).

**E.2 Interfacial thermal conductance between 2-D materials and substrates**

Heat flows across the interfaces between 2-D materials and substrates, and subsequently dissipates to the substrates. The interfacial thermal conductance between graphene or graphite and a wide range of substrates has been studied. The typical values of the interfacial thermal conductance between graphene and substrate materials, such as $SiO_2$, aluminum, and gold thin film, range from 20 to 100 MW/m$^2$K at room temperature and have been summarized in a few articles (Sadeghi, *et al.*, 2012; Xu, *et al.*, 2014b; Ni, *et al.*, 2013). Several groups have measured the interfacial thermal conductance for different thicknesses of graphene, but no systematic dependence on the thickness has been found (Chen, *et al.*, 2009; Mak, *et al.*, 2010).

There exist far fewer measurements on the interfacial thermal conductance of other 2-D materials. Jo *et al*. (Jo, *et al.*, 2013) extracted both thermal conductivity of h-BN and the interfacial thermal conductance (~2MW/m$^2$K) of h-BN on $SiN_x$ substrate from their micro-



bridge experiment. Liu *et al.* (Liu, *et al.*, 2014b) showed that the interfacial thermal conductance between $MoS_2$ and aluminum, NbV and CoPt are 20, 25 and 26 MW/m$^2$K, respectively. These values are lower than the interfacial thermal conductance of Al film on highly ordered pyrolytic graphite (HOPG). Taube *et al.* (Taube, *et al.*, 2015) employed optothermal Raman measurements to extract the interfacial thermal conductance of single-layer $MoS_2$ on $SiO_2$. They found it to be ~1.9 MW/m$^2$K, which is comparable to that of h-BN on $SiN_x$.

There are also several efforts to control the interfacial thermal conductance between 2-D materials and substrates. By corrugating a substrate, the interfacial thermal conductance between graphene and the substrate is reduced by five orders of magnitude relative to a seamless contact (Tang, *et al.*, 2014). Even for perfect interface contact, the interfacial thermal conductance can be actively tuned by applying an electrostatic field, which can improve the conformation between the graphene and substrate (Koh, *et al.*, 2016).

### III. Thermal properties of nanostructures based on 2-D crystals

### A. Defects and alloying

Imperfections, such as point defects, vacancies and grain boundaries, inevitably occur in 2-D materials. These imperfections usually lead to phonon-defect or phonon boundary scattering and thus reduce the thermal conductivity.

Theoretically, the phonon scattering rate due to point defects can be derived from Fermi's golden rule (Tamura, 1983) or be calculated using the atomic Green's functions (Kundu, *et al.*, 2011; Katcho, *et al.*, 2014). For 2-D crystals, the phonon scattering rate can be roughly estimated by (Klemens, 1955; Klemens and Pedraza, 1994; Klemens, 2000):



$$1/\tau^{\mathrm{D}} \propto \left(\omega^{\alpha}/v^{\beta}\right) \sum_i f_i \left[\left(1 - M_i/\bar{M}\right)^2 + \varepsilon\left(\gamma_{\mathrm{G}}\left(1 - r_{a,i}/\bar{r}_a\right)\right)^2\right], \tag{3}$$

where $M_i$, $r_{a,i}$ and $f_i$ are the mass, radius, and fractional concentration of type $i$ atom, $\bar{M}$ and $\bar{r}_a$ are the averaged atom mass and radius, $\gamma_{\mathrm{G}}$ is the Grüneisen parameter, $v$ is the phonon group velocity, and $\varepsilon$ is a phenomenological parameter characterizing the relative importance of defect scattering due to mass difference and lattice distortion. In Eq. (3), $\alpha$ =3 and $\beta$ =2.5 for a linear phonon dispersion and $\alpha = 2$ and $\beta = 1.5$ for a quadratic phonon dispersion, respectively. This expression indicates that the high-frequency (short-wavelength) phonon modes are more likely to be scattered by defects.

Chen *et al*. (Chen*, et al.*, 2012) synthesized graphene sheets with different compositions of carbon isotopes $^{12}$C and $^{13}$C, and measured their thermal conductivity. A small concentration (1.1%) of $^{13}$C isotopes was found to lead to a 30% reduction in thermal conductivity. Pettes *et al*. (Pettes, *et al*., 2015) experimentally studied the effects of isotope scattering on the measured thermal conductivity of high-quality ultrathin graphite samples at both low and high-isotope impurity concentrations. The measured thermal conductivity of the samples was below the value predicted by an incoherent isotope-scattering model derived from perturbation theory, which was attributed to the effects of multiple scattering of phonons by high-concentration isotope impurities in low-dimensional systems. The effect of isotopes was also studied in other 2-D materials, such as h-BN (Lindsay and Broido, 2011b) and MoS$_2$ (Li, *et al.*, 2013), through PBTE calculations. In addition to randomly distributed isotopes, Mingo *et al*. (Mingo*, et al.*, 2010) employed atomistic Green's functions to investigate the possibility of utilizing isotope clusters to tune the thermal conductivity of graphene. As scattering is sensitive to the size of the imperfections, the isotope clusters play a similar role to nanoparticles in bulk materials (Kim*, et*



*al.*, 2006), *i.e.*, scattering low-frequency phonons and thus further reducing the thermal conductivity relative to naturally occurring graphene. The atomistic Green's function calculations show that isotope clustering results in a remarkable increase of the scattering cross-section and thus a significant reduction of the thermal conductivity (Mingo, *et al.*, 2010).

Equation (3) clearly shows that introducing different kinds of atoms could more effectively scatter phonons than the isotopic counterparts, since it introduces both mass differences and lattice distortion. Indeed, lattice distortion in 2-D materials could be quite different from that in 3D bulk materials when other kinds of atoms are introduced. An AGF calculation was carried out to investigate how phonons transmit across a Si defect in graphene (Jiang, *et al.*, 2011a). With small phonon transmission across this defect, thermal conductivity is largely reduced.

Compared to isotopes and substitutions, vacancies can more effectively suppress the thermal conductivity (Hao, *et al.*, 2011; Zhang, *et al.*, 2011). For example, MD simulations by Hao *et al.* (Hao, *et al.*, 2011) showed that the thermal conductivity of graphene with 1% vacancies is reduced to 10-20% of unmodified graphene's, while only a reduction to 70% is found when 1% $^{12}$C atoms are replaced by $^{13}$C atoms (Chen, *et al.*, 2012). Raman-based optothermal measurement of graphene with electron beam irradiation induced defects reveals a thermal conductivity reduction from 1800 W/(m K) to 400 W/(m K) near room temperature as the defect density is slightly changed from $2.0 \times 10^{10}$ cm$^{-2}$ (0.0005%) to $1.8 \times 10^{11}$ cm$^{-2}$ (0.005%) (Malekpour *et al.*, 2016).

From classical perturbation theory, if a vacancy is modeled as if one atom is removed from the crystal along with all the bonding associated with this atom, the effect of vacancy on phonon scattering rate can be estimated by (Ratsifaritana and Klemens, 1987)

$$\frac{1}{\tau_{\mathbf{q}s}^{\mathrm{v}}} \propto \frac{\delta^3}{G_{\mathrm{V}}} \frac{\omega^\alpha}{v^\beta}, \tag{4}$$



where $G_V$ is the number of atoms per defect, $\delta^3$ denotes the atomic volume, and $\alpha$ and $\beta$ have the same meanings as in Eq. (3). Recently, Xie *et al*. (Xie*, et al.*, 2014) pointed out that this classical model neglected the different characteristics of atomic bonds connecting the under-coordinated atoms, which have fewer neighbors to form bonds due to defects than the atoms in perfect crystal.

## B. Heterostructures

Many efforts have been devoted to the synthesis of heterostructures based on a diverse set of 2-D materials. Both in-plane and vertical heterostructures of TMDs have been synthesized (Duan*, et al.*, 2014; Gong*, et al.*, 2014). Thermal transport in such heterostructures has yet to be systematically studied, though there are a few studies on graphene/h-BN and graphene/silicene related structures using MD simulations (Kınacı*, et al.*, 2012; Jing*, et al.*, 2013; Song and Medhekar, 2013; Zhu and Ertekin, 2014; Gao*, et al.*, 2016; Gao*, et al*., 2017) and AGF approach (Jiang*, et al.*, 2011b).

MD simulations showed that for in-plane graphene/h-BN superlattices, the thermal conductivity along the direction of the graphene/h-BN interface is sensitive to the type of interface. If the graphene and h-BN stripes are connected by the zigzag edge, the thermal conductivity is larger than if they are connected by the armchair edge (Sevik*, et al.*, 2011). The effect is more pronounced when the width of the stripe is small, but superlattices with both types of interfaces are found to converge to the average thermal conductivity of graphene and h-BN when the stripes are sufficiently wide. When the heat flux is perpendicular to the interface, the thermal conductivity of the in-plane superlattice of 2-D crystals initially decreases with



deceasing superlattice period, then increases if the period is reduced further (Zhu and Ertekin, 2014). The minimum thermal conductivity was found to be at a critical period of 5 nm.

Thermal transport in graphene with embedded hexagonal h-BN quantum dots has been studied using both a real-space Kubo approach (Sevinçi, *et al.*, 2011), where the phonon transmission is computed through the Kubo formula, and EMD simulations (Sevik, *et al.*, 2011). Both studies showed that the thermal conductivity of graphene with quantum-dots is reduced compared to pristine graphene. The reduction of the thermal conductivity is less distinct with larger quantum dots. The minimal thermal conductivity was found at 50% concentration of h-BN quantum dots with radii of 0.495, 1.238, and 2.476 nm.

Inspired by the high power factor of single-layer $MoS_2$, Gu and Yang (Gu and Yang, 2016) introduce the concept of "nanodomains in 2-D alloy" (which is an analogue of the widely used "nanoparticles in alloy" approach for achieving a high thermoelectric figure of merit in 3-D materials) to reduce the thermal conductivity of 2-D crystals. The $Mo_{0.8}W_{0.2}S_2$ alloy embedded with 10-unit-cell-large triangular $WS_2$ nanodomains was shown to have only 10% of the intrinsic thermal conductivity of single-layer $MoS_2$, due to the strong alloy scattering for high-frequency phonons and the scattering of low-frequency phonons by $WS_2$ nanodomains. It was found that the minimum thermal conductivity alloy is not necessarily the best matrix material for such nanocomposites to achieve the lowest thermal conductivity, since there is a competition between alloy scattering and nano-domain scattering when the alloy is partially replaced with nanodomains. For some TMD alloys, for example $Mo_{1-x}W_xTe_2$, the thermal conductivity could be further tuned by utilizing the composition-dependent phase transition (Yan, *et al.*, 2017; Qian, *et al.*, 2018).



The thermal conductance across interfaces between two different 2-D materials in either vertical or in-plane heterostructures has also been investigated. Chen *et al*. (Chen*, et al.*, 2014) measured the interfacial thermal conductance between vertically stacked graphene and h-BN using the optothermal Raman technique. They found that the interfacial thermal conductance is about 7.4 MW/m$^2$K at room temperature and attributed the measured low interfacial thermal conductance of h-BN/graphene to liquid or organic residues. They believed that the interfacial thermal conductance could be enhanced if the lattice mismatch during graphene growth on h-BN was minimized. MD simulations with a transient heating scheme were also performed to mimic the laser heating in the optothermal Raman measurement. An interfacial thermal conductance of h-BN/graphene was found to be 3.4 MW/m$^2$K, which is consistent with the optothermal Raman measurement (Zhang*, et al.*, 2015a). The simulations also showed that the interfacial thermal conductance could be increased by raising the temperature, increasing the bonding strength or introducing graphene hydrogenation. A similar MD study showed that the thermal conductance of a graphene/MoS$_2$ interface is 5.8 MW/m$^2$K (Liu*, et al.*, 2015). The interfacial thermal conductance of an in-plane interface between graphene and silicene was found to be 250 W/m$^2$K using NEMD simulations (Liu*, et al.*, 2014a).

## C. Surface functionalization and intercalation

Some 2-D materials can be made by chemical functionalization of existing materials. Indeed, many graphene derivatives have been synthesized through hydrogenation, oxidization or fluorination. Compared with single layer pristine graphene, the calculated thermal conductivity of hydrogenated graphene (graphane) and fluorinated graphene (fluorographene) with a 100% coverage of the H and F atoms are reduced to 40%-50% (Pei*, et al.*, 2011; Kim*, et al.*, 2012;



Liu, *et al.*, 2012a) and 35% (Huang, *et al.*, 2012) in MD simulations and ~50% and ~7% in a first-principles-based PBTE study (Fugallo, *et al.*, 2014), respectively. The phonon lifetimes of graphane, extracted from MD simulations, were found to be decreased by an order of magnitude compared with those in pristine graphene (Kim, *et al.*, 2012). The larger scattering rate is due to the coupling between the in-plane and out-of-plane phonon modes (Liu, *et al.*, 2012b). For some other carbon-based 2-D materials, hydrogenation was found to enhance thermal conductivity. This enhancement is thought to be caused by hydrogenation induced anharmonicity reduction from the conversion of $sp^2$ bonding to $sp^3$ bonding.

Unlike in conventional thin film materials, thermal transport in functionalized graphene can also be tuned through the coverage of the functionalized molecules. When the coverage increases from 0% to 100%, the thermal conductivity generally exhibits a U-shape curve (Pei, *et al.*, 2011; Huang, *et al.*, 2012;  Liu, *et al.*, 2012b), similar to the dependence of alloy thermal conductivity on its composition. Mu *et al*. (Mu, *et al.*, 2014) showed that graphene oxide with 20% coverage has a thermal conductivity lower than the calculated thermal conductivity of graphene according to minimum thermal conductivity theory (Cahill, *et al.*, 1992).

Despite the fact that, in most carbon-based materials, the thermal conductivity is reduced after functionalization, Han et al. (Han et al., 2016) showed experimentally that a graphene film deposited on a functionalized graphane could serve as a high-performance heat spreader for electronic chips, which leads to a lower chip temperature than either a graphene film or a graphene film with non-functionalized graphane. Functionalization of graphane not only reduces the contact resistance between graphene and the substrate (chip), but also enhances the thermal conductivity of the graphene film by suppressing the substrate-graphene scattering for flexural phonon modes.



In addition to surface functionalization, intercalation is another facile chemical approach to tune the thermal properties of layered 2-D materials. There has been a rich chemistry in intercalating 2-D materials to form composites or superlattice of layered 2-D materials (Wilson and Yoffe, 1969; Dresselhaus and Dresselhaus, 1981; Whittingha, 2012). Intercalation compounds are formed by the insertion of atomic or molecular layers of a different chemical species (called the intercalant) between the layers of a layered host material, such as graphite, transition metal dichalcogenides, or other *van der Waals* crystals. When the intercalants are inserted into the van der Waals gap of the layered material, charges can be transferred to the layers of the host materials (Wilson and Yoffe, 1969), leading to a change in the Fermi level and thus changing the electron density of states close to the Fermi level, which has tremendous impact on the applications in electronics, photonics, and electrochemical energy strorage. (Dresselhaus and Dresselhaus, 1981)

Besides changing the electronic properties, thermal conductivity also changes significantly from intercalation. Intercalation can both change the magnitude and anisotropy of the thermal conductivity. The study of thermal conductivity of intercalated compounds can be traced back to 1980s (Elzinga*, et al.*, 1982; Issi*, et al.*, 1983; Clarke and Uher, 1984).

To explain the lower lattice thermal conductivity around room temperature, Issi *et al*. (Issi*, et al.*, 1983) employed the PBTE formalism, which considers the scattering of in-plane phonon modes in the intercalation compound's graphene layers (including phonon-phonon scattering, grain boundary scattering, and defect scattering). By fitting the experimental data to the analytical expressions of phonon relaxation times, they attributed the reduced thermal conductivity to the lattice distortion associated with large-scale defects and the grain boundary scattering. However, the extracted feature size of the grains is about one order of magnitude



smaller than their experimental observation, indicating that there are additional scattering mechanisms not considered in their model. Recent MD simulations provided atomic-scale insights into the effect of lithium intercalation on the spectral phonon properties of graphite (Qian, *et al.*, 2016). It was found that the intercalation of lithium ions tunes the anisotropy of graphite's thermal conductivity. The basal plane thermal conductivity experiences a threefold reduction as lithium composition $x$ increases from zero to one in the $Li_xC_6$ (Fig. 11(a)), while the c-axis thermal conductivity decreases from 6.5 W/(m K) (pristine graphite) to 1.8 W/(m K) ($LiC_{18}$), then increases back to 5.0 W/(m K) ($LiC_6$) when fully charged, as shown in Fig. 11(b). Figs 11(c) and 11(d) provides a comparison between pristine graphite and lithiated $LiC_6$. Li ion vibrations are plotted in yellow, while the vibration of the graphite's carbon atoms are plotted in blue. Clearly many low energy flat modes are introduced by lithium intercalation, and these modes gradually merge with the graphite host's vibrational modes. This significantly suppresses the in-plane phonon group velocities, reducing the basal plane thermal conductivity. Such phonon hybridization is also observed in other materials (like clathrates and skutterudites) with rattler atoms weakly coupled to their host material and is found to be responsible for their low thermal conductivity. The non-monotonic change of thermal conductivity along the c-axis direction is explained by two counteractive factors: First, the intercalated lithium ions induce extra electrostatic interaction via the charge transfer process. This enhanced inter-layer coupling increases the phonon velocities along the c-axis relative to pristine graphite, as shown in Figs. 11(e) and 11(f). Second, the lithium ions also provide extra phonon scattering channels, suppressing the phonon lifetime. These two opposing factors together render the non-monotonic change in the thermal conductivity along the c-axis.



Although many other intercalation compounds aside from graphite intercalation compounds have been synthesized, there are only limited thermal conductivity measurements. $TiS_2$-based intercalation compounds have attracted considerable attention due to the large thermoelectric power of bulk $TiS_2$ crystals (Imai, *et al.*, 2001). Wan *et al.* (Wan, *et al.*, 2010a; Wan, *et al.*, 2010b; 2011; Wan, *et al.*, 2012) inserted different intercalants, such as **SnS and BiS**, into the $TiS_2$ van der Waals gap to form natural superlattices and demonstrated an improved thermoelectric figure of merit (ZT)`. One example is the superlattice $(SnS)_1(TiS_2)_2$, where SnS layer is intercalated into $TiS_2$ layered crystal. The cross-plane thermal conductivity is shown to be lower than the prediction of the minimum thermal conductivity theory. The authors attributed this low thermal conductivity to phonon localization due to the orientation disorder of the SnS layers (Wan, *et al.*, 2011).

Organic molecules can also be inserted into the van der Waals gaps. Wan et al. recently synthesized $TiS_2/[(Hexylammonium)_{0.08}(H2O)_{0.22}(DMSO)_{0.03}]$ (an inorganic/organic layered hybrid material) for thermoelectric applications (Wan, *et al.*, 2015a; Wan, *et al.*, 2015b) (Fig. 12(a)). Due to its low lattice thermal conductivity (Fig. 12(b)) and relatively unchanged power factor from $TiS_2$ crystals, the thermoelectric figure of merit ZT is measured to be as high as 0.42, making the hybrid organic-inorganic superlattice a promising n-type flexible thermoelectric material (Wan, *et al.*, 2015a). The measured thermal conductivity of the hybrid material is only one-sixth of that of the bulk counterpart. More surprisingly, the lattice thermal conductivity was only 1/30 of bulk $TiS_2$ after using the Weidmann-Franz law to extract the electronic contribution. MD simulations were performed to calculate the thermal conductivity of bulk $TiS_2$ and hybrid organic-inorganic materials and an eight-fold reduction of lattice thermal conductivity was found for $TiS_2$ when intercalated with organic components. This reduction is significantly higher than



simple volumetric averaging, indicating that the inorganic-organic coupling plays an important role in in-plane phonon transport.

The thermal conductivity of lithium-intercalated compounds of cobalt oxide (Cho, *et al.*, 2014) and $MoS_2$ (Zhu, *et al.*, 2016a) were recently measured using the transient thermoreflectance technique. The thermal conductivity of both compounds was found to be tunable with the amount of Li ions intercalated. In particular, intercalation of lithium changes the thermal conductivity anisotropy in $MoS_2$ (Zhu, *et al.*, 2016a). The basal-plane thermal conductivity decreases monotonically but the cross-plane thermal conductivity decreases and then increases with increasing lithium composition. This experimental trend was found to be similar to MD simulation results for the lithium intercalated graphite compounds (Qian, *et al.*, 2016).

## IV.  Summary and Outlook

### A.     Summary

Through a comprehensive survey of a large body of literatures, we have answered most of the questions on phonon and thermal properties in 2-D materials raised at beginning of this Colloquium:

a) In general, the phononic thermal conductivity of a suspended 2-D material keeps increasing with the sample size even when the size is beyond tens of microns. Some first principles calculations suggested that it converges when the sample is about 1 mm. However, this has not yet been experimentally confirmed. Strain and crystal structure can significantly affect the scattering of long-wavelength acoustic phonons, complicating the thermal conductivity's length dependence.



b) For suspended 2-D materials, thermal conductivity decreases with the increasing number of layers because inter-layer interactions suppress the flexural modes. For some 2-D crystals, the thickness dependence of thermal conductivity can exhibit a different trend due to the distinct crystal structures and the corresponding phonon dispersion and scattering properties.

c) The thermal conductivity of any supported 2-D materials is a finite value because of the suppression of flexural modes and the breakdown of the translational invariance. The thermal conductivity of 2-D materials and the interfacial thermal conductance can be modulated by a substrate. When the structure of the substrate is very similar to that of 2-D materials, the reduction is minimized. Mechanical strain plays an important role in modulating thermal conductivity.

d) Defects and functionalization can considerably reduce thermal conductivity of 2-D materials. Intercalation could affect both the group velocities and phonon relaxation times of layered crystals, thus tuning the thermal conductivity along the through-plane and basal-plane directions.

Many remarkable achievements and progresses have been made in the past decade on phonon and thermal properties of 2-D materials including graphene, however; there are still many challenging theoretical and experimental questions that need to be solved.

Theoretically, there is no rigorous mathematical proof on whether there exists size dependence of intrinsic thermal conductivity of 2-D materials. The modeling of 2-D lattices has been limited to the in-plane motion of lattices (Lippi and Livi, 2000; Yang, *et al.*, 2006; Xiong, *et al.*, 2010; Wang, *et al.*, 2012a). This simplification is used because theoretical analysis (e.g.



mode-mode coupling) becomes extremely difficult when flexural modes are considered. However, in order to understand thermal transport in realistic 2-D materials, the flexural motion should not be neglected. In the meantime, significant progresses have been made in the study of heat conduction in 1-D system over the last few decades. With the great efforts from both mathematicians and theoretical physicists, we now know that thermal conductivity in 1-D momentum conserved systems can exhibit anomalous behavior, although more experimental works are still needed to verify this theoretical predication. We hope that this summary of the anomalous heat conduction behavior in 2-D systems and 2-D materials (even if it is only up to a certain length) in this Colloquium can help attract attentions from theoreticians and experimentalists.

Experimentally, only limited techniques are available for measuring thermal conductivity of 2-D materials. The optothermal Raman spectroscopy that was used to obtain the first experimental data on thermal conductivity of suspended single layer graphene, has been found to overestimate thermal conductivity. This method relies on two conditions: (1) absorption of laser energy by the materials, and (2) the temperature dependence of the phonon peak shift. Any wrong calculation of these two factors could result in a significant systematic error. The first one has been known for many years, however, the second one has only been investigated very recently by Vallabhaneni *et al*. (Vallabhaneni, *et al.*, 2016). In their detailed computational study, Vallabhaneni *et al.* pointed out that the ZA and ZO phonons that contribute the most to the thermal conductivity of single layer graphene are indeed out of thermal equilibrium. The temperature of these phonon modes is found to be much lower than the temperature measured from the shift of Raman peak frequency. The latter is assumed to be the equilibrium lattice temperature. This leads to an overestimate of the thermal conductivity by a factor of 1.35 ~ 2.6 at



room temperature. How to incorporate such theoretical understanding into optothermal Raman spectroscopy to obtain a more accurate measurement of thermal conductivity is an interesting new direction (Sullivan, *et al.*, 2017).

## B. Outlook

### Dimensional crossover from 2-D to 1-D, and from 2-D to 3-D

2-D materials provide an ideal test platform to study the dimensional crossover of phonon transport. When studying size effect of 2-D materials, in principle we simultaneously change both the width and length of samples to preserve the two-dimensionality. However, in experiments it is often more convenient to keep the width fixed and only the length is changed. Indeed this could be an effective way to study the dimensional crossover from 2-D to 1-D. A transition signature was reported in a recent experiment on thermal conductivity versus length for a suspended single layer graphene (Xu, *et al.*, 2014a). It could also be interesting to know how many layers are necessary to render a bulk thermal conductivity of layered materials.

### Electronic thermal transport

Eelectrons also carry heat in both metals and semiconductors. In bulk materials, the electronic thermal conductivity and electrical conductivity are related by the well-known Wiedemann-Franz law (Kittel, 2005). However, in low dimensional systems, as we have discussed in this Colloquium article, heat conduction becomes anomalous and it is not clear if the Wiedemann-Franz law is still valid. Lee *et al.* recently found that at 240 – 340 K the Wiedemann-Franz law is broken in metallic vanadium dioxide ($VO_2$) (Lee, *et al.*, 2017a). There is no doubt that there might also be anomalous thermal transport for electrons in 2-D materials.



**Electron-phonon coupling**

It is known that in both the Raman optothermal method and other laser heating/probing methods, the heating energy is transferred first from photons to electrons, and then from electrons to phonons through electron-phonon coupling. The electron-phonon coupling determines how much energy can be transferred from hot electrons to phonons. The coupling also determines the temperature of the phonons, if non-equilibrium transport is neglected by assuming a sufficiently long time window in thermal metrology. A better understanding of electron-phonon coupling in 2D materials helps to improve the measurement accuracy of thermal conductivity.

**Topological phonon effect in 2-D materials**

2-D materials provide a test platform for topological phonon effect. The phonon Hall effect was discovered (Strohm*, et al.*, 2005; Inyushkin and Taldenkov, 2007) in crystals of paramagnetic terbium gallium garnet (TGG), Tb3Ga5O12. These cubic crystals are dielectrics and contain ions that carry both a high charge and a large magnetic moment; both factors may induce a strong coupling to an applied magnetic field, leading to the phonon Hall effect.

A theoretical study by Zhang *et al.* (Zhang*, et al.*, 2010) shows that on a 2-D honeycomb lattice the phonon Hall effect arises due to the Berry phase induced by the band curvature. Moreover, this phase is quantized in the multiples of the Chern number indicating that the phonon Hall effect is intrinsically a topological effect. Additional analysis by Qin *et al.* (Qin*, et al.*, 2012) confirmed that the thermal conductivity in the Hall regime is the same as that of a series of 1-D quantum wires, which is consistent with the expectation that a topological material



develops nontrivial edge modes.

The phonon Hall effect remains unexplored in many realistic 2-D materials, both theoretically and experimentally. In particular, in magnetically ordered materials like yttrium iron garnet (YIG), the presence of strong magnetoelastic spin-phonon couplings allows the hybridization of magnons and phonons.

## Phononic Quantum Devices with 2-D materials

The novel characteristics of phonons in 2-D materials can find applications in an emerging field of quantum engineering. Because of their extremely low mass, 2-D materials have been used as a nanoscale membrane to build a nanoscale drum (optomechanical cavity), which enables quantum information storage. Moreover, 2-D materials can be used as opto-mechanical sensors with unprecedented sensitivities. For example, Singh *et al.* (2014) have demonstrated a displacement sensitivity of 17 fm/Hz$^{1/2}$ when a multilayer graphene resonator, with Q = 220,000, is coupled to a high Q superconducting cavity. The results show a potential for phonon storage for up to 10 ms, a significant improvement compared with other mechanical resonators like silicon microsphere (3 microsecond) (Fiore*, et al.*, 2011), and superconducting aluminum membrane (90 microsecond) (Palomaki*, et al.*, 2013). In addition to the quantum storage demonstrated by Singh *et al.* (2014), propagating phonons can be also used to transmit quantum information when coupled to artificial atoms in the quantum regime (Gustafsson*, et al.*, 2014). These works have opened a new direction of phonons in quantum information and computation (Aspelmeyer*, et al.*, 2014; Tian, 2015).

## Anomalous thermal conductivity and the second sound



Another interesting topic that worth exploring is whether the anomalous thermal conductivity in 2D materials has any connection with the second sound – a transport of heat in the form of a wave. The second sound is a spatially periodic fluctuation of entropy and temperature. This is different from conventional sound – acoustic wave- that is a periodic fluctuation of mass density and pressure. The second sound concept was proposed (Donnelly, 2009) for Helium II where both the normal fluid and superfluid co-exist, and has been experimentally observed in hydrodynamic materials such as solid Helium (Ackerman*, et al.*, 1966), NaF (Jackson*, et al.*, 1970; Pohl and Irniger, 1976), bismuth (Narayanamurti and Dynes, 1972), SrTiO$_3$ (Koreeda*, et al.*, 2007) at very low temperature. A theoretical question one would naturally ask is whether the second sound exists in 2-D materials at room temperature. There have been some theoretical studies along this direction (Cepellotti*, et al.*, 2015; Lee*, et al.*, 2015a). It would be very interesting to study this experimentally if the speed of the second sound can be probed.


**Acknowledgement:** BL would like to thank J. Chen, D. Donadio, B. Hu, S.-Q. Hu, W. Li, D. Liu, B. Özyilmaz, J. Ren, J.T.L. Thong, J.-S. Wang, J.-Y. Wang, L. Wang, R.-G. Xie, X.-F. Xu, N. Yang, L.-F. Zhang, Z.-W. Zhang, L.-Y. Zhu, and G. Zhang for fruitful collaborations in 2-D systems in past years. RY would like to thank M.S. Dresselhaus, X.B. Li, J. Liu, J. Zhu, X.J. Wang, Y.K. Koh, C.L. Wan, K. Koumoto, P.Q. Jiang, and X. Qian, for their collaborative work on thermal transport in 2-D materials over the past 10 years or so. The funding support for RY was provided by National Science Foundation (Grant No. 0846561, 1512776) and DARPA (Grant No. FA8650-15-1-7524). YW acknowledges the support from National Natural Science Foundation of China (Grant No. 11425211) and the Strategic Priority Research Program of the




Chinese Academy of Sciences (XDB22020200). We thank S.R. Sklan for reviewing and editing the final version of this article.

**References:**

Ackerman, C. C., B. Bertman, H. A. Fairbank, and R. Guyer, 1966, "Second sound in solid helium," Physical Review Letters **16**, 789.

Aksamija, Z., and I. Knezevic, 2011, "Lattice thermal conductivity of graphene nanoribbons: Anisotropy and edge roughness scattering," Applied Physics Letters **98**, 141919.

Alam, M., R. Pulavarthy, C. Muratore, and M. A. Haque, 2015, "Mechanical strain dependence of thermal transport in amorphous silicon thin films," Nanoscale and Microscale Thermophysical Engineering **19**, 1.

Ashcroft, N. W., and N. D. Mermin, 1978, Solids State Physics (Holt, Rinehart and Winston).

Aspelmeyer, M., T. J. Kippenberg, and F. Marquardt, 2014, "Cavity optomechanics," Reviews of Modern Physics **86**, 1391.

Bae, M. H., Z. Li, Z. Aksamija, P. N. Martin, F. Xiong, Z. Y. Ong, I. Knezevic, and E. Pop, 2013, "Ballistic to diffusive crossover of heat flow in graphene ribbons," Nature Communications **4**, 1734.

Balandin, A. A., 2011, "Thermal properties of graphene and nanostructured carbon materials," Nature Materials **10**, 569.

Balandin, A. A., S. Ghosh, W. Bao, I. Calizo, D. Teweldebrhan, F. Miao, and C. N. Lau, 2008, "Superior thermal conductivity of single-layer graphene," Nano Letters **8**, 902.




Barbarino, G., C. Melis, and L. Colombo, 2015, "Intrinsic thermal conductivity in monolayer graphene is ultimately upper limited: A direct estimation by atomistic simulations," Physical Review B **91**, 035416.

Berciaud, S., S. Ryu, L. E. Brus, and T. F. Heinz, 2008, "Probing the intrinsic properties of exfoliated graphene: Raman spectroscopy of free-standing monolayers," Nano Letters **9**, 346.

Bhowmick, S., and V. B. Shenoy, 2006, "Effect of strain on the thermal conductivity of solids," The Journal of Chemical Physics **125**, 164513.

Bonini, N., J. Garg, and N. Marzari, 2012, "Acoustic phonon lifetimes and thermal transport in free-standing and strained graphene," Nano Letters **12**, 2673.

Broido, D., M. Malorny, G. Birner, N. Mingo, and D. Stewart, 2007, "Intrinsic lattice thermal conductivity of semiconductors from first principles," Applied Physics Letters **91**, 231922.

Broido, D., A. Ward, and N. Mingo, 2005, "Lattice thermal conductivity of silicon from empirical interatomic potentials," Physical Review B **72**, 014308.

Cahill, D. G., S. K. Watson, and R. O. Pohl, 1992, "Lower limit to the thermal conductivity of disordered crystals," Physical Review B **46**, 6131.

Cai, W., A. L. Moore, Y. Zhu, X. Li, S. Chen, L. Shi, and R. S. Ruoff, 2010, "Thermal transport in suspended and supported monolayer graphene grown by chemical vapor deposition," Nano Letters **10**, 1645.

Cao, H.-Y., Z.-X. Guo, H. Xiang, and X.-G. Gong, 2012, "Layer and size dependence of thermal conductivity in multilayer graphene nanoribbons," Physics Letters A **376**, 525.





Carrete, J., W. Li, L. Lindsay, D. A. Broido, L. J. Gallego, and N. Mingo, 2016, "Physically founded phonon dispersions of few-layer materials and the case of borophene," Materials Research Letters **4**, 204.

Castellanos-Gomez, A., L. Vicarelli, E. Prada, J. O. Island, K. Narasimha-Acharya, S. I. Blanter, D. J. Groenendijk, M. Buscema, G. A. Steele, and J. Alvarez, 2014, "Isolation and characterization of few-layer black phosphorus," 2D Materials **1**, 025001.

Cepellotti, A., G. Fugallo, L. Paulatto, M. Lazzeri, F. Mauri, and N. Marzari, 2015, "Phonon hydrodynamics in two-dimensional materials," Nature Communications **6**, 6400.

Chang, C.-W., 2016, in *Experimental Probing of Non-Fourier Thermal Conductors,* edited by Lepri, S. (Springer International Publishing), p. 305.

Chen, C.-C., Z. Li, L. Shi, and S. B. Cronin, 2014, "Thermal interface conductance across a graphene/hexagonal boron nitride heterojunction," Applied Physics Letters **104**, 081908.

Chen, G., 2001, "Ballistic-diffusive heat-conduction equations," Physical Review Letters **86**, 2297.

Chen, G., 2005, *Nanoscale Energy Transport and Conversion: A Parallel Treatment of Electrons, Molecules, Phonons, and Photons* (Oxford University Press, New York).

Chen, J., S. Chen, and Y. Gao, 2016, "Anisotropy Enhancement of Thermal Energy Transport in Supported Black Phosphorene," The Journal of Physical Chemistry Letters **7**, 2518.

Chen, J., G. Zhang, and B. Li, 2013, "Substrate coupling suppresses size dependence of thermal conductivity in supported graphene," Nanoscale **5**, 532.

Chen, L., and S. Kumar, 2012, "Thermal transport in graphene supported on copper," Journal of Applied Physics **112**, 043502.





Chen, S., A. L. Moore, W. Cai, J. W. Suk, J. An, C. Mishra, C. Amos, C. W. Magnuson, J. Kang, and L. Shi, 2010, "Raman measurements of thermal transport in suspended monolayer graphene of variable sizes in vacuum and gaseous environments," ACS Nano **5**, 321.

Chen, S., J. Wang, G. Casati, and G. Benetti, 2014, "Nonintegrability and the Fourier heat conduction law," Physical Review E **90**, 032134.

Chen, S., Q. Wu, C. Mishra, J. Kang, H. Zhang, K. Cho, W. Cai, A. A. Balandin, and R. S. Ruoff, 2012, "Thermal conductivity of isotopically modified graphene," Nature Materials **11**, 203.

Chen, S., Y. Zhang, J. Wang, and H. Zhao, 2016, "Key role of asymmetric interactions in low-dimensional heat transport", Journal of Statistical Mechanics: Theory and Experiment 3, 033205.

Chen, Y., J. Analytis, J.-H. Chu, Z. Liu, S.-K. Mo, X.-L. Qi, H. Zhang, D. Lu, X. Dai, and Z. Fang, 2009, "Experimental realization of a three-dimensional topological insulator, Bi2Te3," Science **325**, 178.

Chen, Z., W. Jang, W. Bao, C. N. Lau, and C. Dames, 2009, "Thermal contact resistance between graphene and silicon dioxide," Applied Physics Letters, **95**, 161910.

Cho, J., M. D. Losego, H. G. Zhang, H. Kim, J. Zuo, I. Petrov, D. G. Cahill, and P. V. Braun, 2014, "Electrochemically tunable thermal conductivity of lithium cobalt oxide," Nature Communications **5**, 4035.

Clarke, R., and C. Uher, 1984, "High pressure properties of graphite and its intercalation compounds," Advances in Physics **33**, 469.





Cocemasov, A. I., D. L. Nika, and A. A. Balandin, 2015, "Engineering of the thermodynamic properties of bilayer graphene by atomic plane rotations: the role of the out-of-plane phonons," Nanoscale **7**, 12851.

Conley H. J., B. Wang, J. I. Ziegler, R. F. Haglund Jr., S. T. Pantelides, and K. I. Bolotin, 2013 Bandgap Engineering of Strained Monolayer and Bilayer $MoS_2$, Nano Letters **13**, 3626.

Das, S. G., A. Dhar, and O. Narayan, 2014, "Heat Conduction in the $\alpha - \beta$ Fermi–Pasta–Ulam Chain," Journal of Statistical Physics **154**, 204.

Dickey, J., and A. Paskin, 1969, "Computer simulation of the lattice dynamics of solids," Physical Review **188**, 1407.

Ding, Z., Q.-X. Pei, J.-W. Jiang, and Y.-W. Zhang, 2015, "Manipulating the thermal conductivity of monolayer MoS2 via lattice defect and strain engineering," The Journal of Physical Chemistry C **119**, 16358.

Donnelly, R. J., 2009, "The two-fluid theory and second sound in liquid helium," Physics Today **62**, 34.

Dresselhaus, M. S., and G. Dresselhaus, 1981, "Intercalation compounds of graphite," Advances in Physics **30**, 139.

Duan, X., C. Wang, J. C. Shaw, R. Cheng, Y. Chen, H. Li, X. Wu, Y. Tang, Q. Zhang, A. Pan, J. Jiang, R. Yu, Y. Huang, and X. Duan, 2014, "Lateral epitaxial growth of two-dimensional layered semiconductor heterojunctions," Nature Nanotechnology **9**, 1024.

Ecsedy, D. J., and P. G. Klemens, 1977, "Thermal resistivity of dielectric crystals due to four-phonon processes and optical modes," Physical Review B **15**, 5957.

Elzinga, M., D. Morelli, and C. Uher, 1982, "Thermal transport properties of $SbCl_5$ graphite," Physical Review B **26**, 3312.





Ernst, M., E. Hauge, and J. Van Leeuwen, 1971, "Asymptotic time behavior of correlation functions. I. Kinetic terms," Physical Review A **4**, 2055.

Ernst, M., E. Hauge, and J. Van Leeuwen, 1976a, "Asymptotic time behavior of correlation functions. II. Kinetic and potential terms," Journal of Statistical Physics **15**, 7.

Ernst, M., E. Hauge, and J. Van Leeuwen, 1976b, "Asymptotic time behavior of correlation functions. III. Local equilibrium and mode-coupling theory," Journal of Statistical Physics **15**, 23.

Esfarjani, K., G. Chen, and H. T. Stokes, 2011, "Heat transport in silicon from first-principles calculations," Physical Review B **84**, 085204.

Evans, W. J., L. Hu, and P. Keblinski, 2010, "Thermal conductivity of graphene ribbons from equilibrium molecular dynamics: Effect of ribbon width, edge roughness, and hydrogen termination," Applied Physics Letters **96**, 203112.

Fan, Z., L. F. C. Pereira, P. Hirvonen, M. M. Ervasti, K.R. Elder, D. Donadio, T. Ala-Nissila, and A. Harju, 2017, "Thermal conductivity decomposition in two-dimensional materials: Application to graphene," Physical Review B 95, 144309.

Fasolino, A., J. H. Los, and M. I., Katsnelson, 2007, "Intrinsic ripples in grapheme," Nature Materials **6**, 858.

Faugeras, C., B. Faugeras, M. Orlita, M. Potemski, R. R. Nair, and A. Geim, 2010, "Thermal conductivity of graphene in corbino membrane geometry," ACS Nano **4**, 1889.

Feng T, L Lindsay, X. L Ruan 2017 "Four-phonon scattering significantly reduces intrinsic thermal conductivity of solids," Physical Review B **96**, 161201(R).





Ferrari, A., J. Meyer, V. Scardaci, C. Casiraghi, M. Lazzeri, F. Mauri, S. Piscanec, D. Jiang, K. Novoselov, and S. Roth, 2006, "Raman spectrum of graphene and graphene layers," Physical Review Letters **97**, 187401.

Fiore, V., Y. Yang, M. C. Kuzyk, R. Barbour, L. Tian, and H. Wang, 2011, "Storing optical information as a mechanical excitation in a silica optomechanical resonator," Physical Review Letters **107**, 133601.

Fu, Q., J. Yang, Y. Chen, D. Li, and D. Xu, 2015, "Experimental evidence of very long intrinsic phonon mean free path along the c-axis of graphite," Applied Physics Letters **106**, 031905.

Fugallo, G., A. Cepellotti, L. Paulatto, M. Lazzeri, N. Marzari, and F. Mauri, 2014, "Thermal conductivity of graphene and graphite: collective excitations and mean free paths," Nano Letters **14**, 6109.

Fugallo, G., M. Lazzeri, L. Paulatto, and F. Mauri, 2013, "Ab initio variational approach for evaluating lattice thermal conductivity," Physical Review B **88**, 045430.

Gandi, A. N., H. N. Alshareef, and U. Schwingenschlögl, 2016, "Thermal response in van der Waals heterostructures," Journal of Physics: Condensed Matter **29**, 035504.

Gao Y,. Q.-C. Liu, and B.-X Xu, 2016, "Lattice Mismatch Dominant Yet Mechanically Tunable Thermal Conductivity in Bilayer Heterostructures," ACS Nano, **10**, 5431.

Gao Y., W. Z. Yang, and B.-X. Xu, 2017, "Tailoring Auxetic and Contractile Graphene to Achieve Interface Structures with Fully Mechanically Controllabl Thermal Transports," Advanced Materials Interfaces **10**, 1700278.



Garg, J., N. Bonini, B. Kozinsky, and N. Marzari, 2011, "Role of disorder and anharmonicity in the thermal conductivity of silicon-germanium alloys: A first-principles study," Physical Review Letters **106**, 045901.

Gendelman, O.V., and A. V. Savin, 2000, "Normal Heat Conductivity of the One-Dimensional Lattice with Periodic Potential of Nearest-Neighbor Interaction," Physical Review Letters 84, 2381.

Gendelman, O.V., and A. V. Savin, 2014, "Normal heat conductivity in chains capable of dissociation," EPL 106, 34004.

Ghosh, S., W. Bao, D. L. Nika, S. Subrina, E. P. Pokatilov, C. N. Lau, and A. A. Balandin, 2010, "Dimensional crossover of thermal transport in few-layer graphene," Nature Materials **9**, 555.

Giardina, C. R. Livi, A. Politi, and M. Vassalli, 2000, "Finite Thermal Conductivity in 1D Lattices," Physical Review Letters 84, 2144

Gong, Y., J. Lin, X. Wang, G. Shi, S. Lei, Z. Lin, X. Zou, G. Ye, R. Vajtai, and B. I. Yakobson, 2014, "Vertical and in-plane heterostructures from $WS_2/MoS_2$ monolayers," Nature Materials **13**, 1135.

Goyal, V., D. Teweldebrhan, and A. A. Balandin, 2010, "Mechanically-Exfoliated Stacks of Thin Films of Bismuth Telluride Topological Insulators with Enhanced Thermoelectric Performance," Applied Physics Letters, **97**, 133117.

Gu, X., B. Li, and R.G. Yang, 2016, "Layer thickness-dependent phonon properties and thermal conductivity of MoS2," Journal of Applied Physics **119**, 085106.

Gu, X., and R.G. Yang, 2014, "Phonon transport in single-layer transition metal dichalcogenides: A first-principles study," Applied Physics Letters **105**, 131903.





Gu, X., and R.G. Yang, 2015, "First-principles prediction of phononic thermal conductivity of silicene: A comparison with graphene," Journal of Applied Physics **117**, 025102.

Gu, X., and R.G. Yang, 2016, "Phonon transport in single-layer M o 1− x W x S 2 alloy embedded with W S 2 nanodomains," Physical Review B **94**, 075308.

Guo, Z., D. Zhang, and X.-G. Gong, 2009, "Thermal conductivity of graphene nanoribbons," Applied Physics Letters **95**, 163103.

Gustafsson, M. V., T. Aref, A. F. Kockum, M. K. Ekström, G. Johansson, and P. Delsing, 2014, "Propagating phonons coupled to an artificial atom," Science **346**, 207.

Han H., Y. Zhang, N. Wang, M. K. Samani, Y. Ni, Z. Y. Mijbil, M. Edwards, S. Xiong, K. Sääskilahti, M. Murugesan, Y. Fu, L. Ye, H. Sadeghi, S. Bailey, Y. A. Kosevich, C. J. Lambert, J. Liu, and S. Volz, 2016, "Functionalization mediates heat transport in graphene nanoflakes," Nature Communications **7**, 11281.

Hao, F., D. Fang, and Z. Xu, 2011, "Mechanical and thermal transport properties of graphene with defects," Applied Physics Letters **99**, 041901.

Haskins, J., A. Kınacı, C. Sevik, H. l. Sevinçli, G. Cuniberti, and T. Çağın, 2011, "Control of thermal and electronic transport in defect-engineered graphene nanoribbons," ACS Nano **5**, 3779.

Henry, A., and G. Chen, 2008, "High thermal conductivity of single polyethylene chains using molecular dynamics simulations," Physical Review Letters **101**, 235502.

Henry, A., and G. Chen, 2009, "Anomalous heat conduction in polyethylene chains: Theory and molecular dynamics simulations," Physical Review B **79**, 144305.





Hossain, M. S., F. Al-Dirini, F. M. Hossain, and E. Skafidas, 2015, "High performance graphene nano-ribbon thermoelectric devices by incorporation and dimensional tuning of nanopores," Scientific Reports **5**, 11297.

Hsieh, W.-P., B. Chen, J. Li, P. Keblinski, and D. G. Cahill, 2009, "Pressure tuning of the thermal conductivity of the layered muscovite crystal," Physical Review B **80**, 180302.

Hsieh, W.-P., M. D. Losego, P. V. Braun, S. Shenogin, P. Keblinski, and D. G. Cahill, 2011, "Testing the minimum thermal conductivity model for amorphous polymers using high pressure," Physical Review B **83**, 174205.

Hu, J., X. Ruan, and Y. P. Chen, 2009, "Thermal conductivity and thermal rectification in graphene nanoribbons: a molecular dynamics study," Nano Letters **9**, 2730.

Hu, J., S. Schiffli, A. Vallabhaneni, X. Ruan, and Y. P. Chen, 2010, "Tuning the thermal conductivity of graphene nanoribbons by edge passivation and isotope engineering: A molecular dynamics study," Applied Physics Letters **97**, 133107.

Huang, W., Q.-X. Pei, Z. Liu, and Y.-W. Zhang, 2012, "Thermal conductivity of fluorinated graphene: A non-equilibrium molecular dynamics study," Chemical Physics Letters **552**, 97.

Imai, H., Y. Shimakawa, and Y. Kubo, 2001, "Large thermoelectric power factor in TiS 2 crystal with nearly stoichiometric composition," Physical Review B **64**, 241104.

Inyushkin, A. V. e., and A. Taldenkov, 2007, "On the phonon Hall effect in a paramagnetic dielectric," JETP Letters **86**, 379.

Issi, J.-P., J. Heremans, and M. S. Dresselhaus, 1983, "Electronic and lattice contributions to the thermal conductivity of graphite intercalation compounds," Physical Review B **27**, 1333.





Jackson, H. E., C. T. Walker, and T. F. McNelly, 1970, "Second sound in NaF," Physical Review Letters **25**, 26.

Jain, A., and A. J. McGaughey, 2015, "Strongly anisotropic in-plane thermal transport in single-layer black phosphorene," Scientific Reports **5**, 8501.

Jang, H., J. D. Wood, C. R. Ryder, M. C. Hersam, and D. G. Cahill, 2015, "Anisotropic thermal conductivity of exfoliated black phosphorus," Advanced Materials **27**, 8017.

Jang, W., W. Bao, L. Jing, C. Lau, and C. Dames, 2013, "Thermal conductivity of suspended few-layer graphene by a modified T-bridge method," Applied Physics Letters **103**, 133102.

Jiang, J.-W., H. S. Park, and T. Rabczuk, 2013, "Molecular dynamics simulations of single-layer molybdenum disulphide (MoS2): Stillinger-Weber parametrization, mechanical properties, and thermal conductivity," Journal of Applied Physics **114**, 064307.

Jiang, J.-W., B.-S. Wang, and J.-S. Wang, 2011a, "First principle study of the thermal conductance in graphene nanoribbon with vacancy and substitutional silicon defects," Applied Physics Letters **98**, 113114.

Jiang, J.-W., J.-S. Wang, and B.-S. Wang, 2011b, "Minimum thermal conductance in graphene and boron nitride superlattice," Applied Physics Letters **99**, 043109.

Jo, I., M. T. Pettes, J. Kim, K. Watanabe, T. Taniguchi, Z. Yao, and L. Shi, 2013, "Thermal conductivity and phonon transport in suspended few-layer hexagonal boron nitride," Nano Letters **13**, 550.

Jo, I., M. T. Pettes, E. Ou, W. Wu, and L. Shi, 2014, "Basal-plane thermal conductivity of few-layer molybdenum disulfide," Applied Physics Letters **104**, 201902.





Ju, Y., and K. Goodson, 1999, "Phonon scattering in silicon films with thickness of order 100 nm," Applied Physics Letters **74**, 3005.

Katcho, N., J. Carrete, W. Li, and N. Mingo, 2014, "Effect of nitrogen and vacancy defects on the thermal conductivity of diamond: An ab initio Green's function approach," Physical Review B **90**, 094117.

Kim, J. Y., J.-H. Lee, and J. C. Grossman, 2012, "Thermal transport in functionalized graphene," ACS Nano **6**, 9050.

Kim, P., L. Shi, A. Majumdar, and P. L. McEuen 2001, "Thermal Transport Measurements of Individual Multiwalled Nanotubes," Physcical Review Letters **87**,, 215502.

Kim, W., J. Zide, A. Gossard, D. Klenov, S. Stemmer, A. Shakouri, and A. Majumdar, 2006, "Thermal conductivity reduction and thermoelectric figure of merit increase by embedding nanoparticles in crystalline semiconductors," Physical Review Letters **96**, 045901.

Kınacı, A., J. B. Haskins, C. Sevik, and T. Çağın, 2012, "Thermal conductivity of BN-C nanostructures," Physical Review B **86**, 115410.

Kittel, C., 2005, *Introduction to Solid State Physics* (Wiley, Hoboken, NJ).

Klemens, P. G., 1955, "The scattering of low-frequency lattice waves by static imperfections," Proceedings of the Physical Society. Section A **68**, 1113.

Klemens, P. G., 2000, "Theory of the a-plane thermal conductivity of graphite," Journal of Wide Bandgap Materials **7**, 332.

Klemens, P. G., and D. F. Pedraza, 1994, "Thermal conductivity of graphite in the basal plane," Carbon **32**, 735.





Koh, Y. K., A. S. Lyons, M.-H. Bae, B. Huang, V. E. Dorgan, D. G. Cahill, and E. Pop, 2016, "Role of remote interfacial phonon (RIP) scattering in heat transport across graphene/SiO2 interfaces," Nano Letters **16**, 6014.

Koreeda, A., R. Takano, and S. Saikan, 2007, "Second sound in SrTiO 3," Physical Review Letters **99**, 265502.

Kuang, Y., L. Lindsay, and B. Huang, 2015, "Unusual enhancement in intrinsic thermal conductivity of multilayer graphene by tensile strains," Nano Letters **15**, 6121.

Kuang, Y., L. Lindsay, S. Shi, X. Wang, and B. Huang, 2016a, "Thermal conductivity of graphene mediated by strain and size," International Journal of Heat and Mass Transfer **101**, 772.

Kuang, Y. D., L. Lindsay, S. Q. Shi, and G. Zheng, 2016b, "Tensile strains give rise to strong size effects for thermal conductivities of silicene, germanene and stanene," Nanoscale **8**, 3760.

Kundu, A., N. Mingo, D. Broido, and D. Stewart, 2011, "Role of light and heavy embedded nanoparticles on the thermal conductivity of SiGe alloys," Physical Review B **84**, 125426.

Lee, J.-U., D. Yoon, H. Kim, S. W. Lee, and H. Cheong, 2011, "Thermal conductivity of suspended pristine graphene measured by Raman spectroscopy," Physical Review B **83**, 081419.

Lee, S., D. Broido, K. Esfarjani, and G. Chen, 2015a, "Hydrodynamic phonon transport in suspended graphene," Nature Communications **6**, 6290.

Lee, S., K. Hippalgaonkar, F. Yang, J. Hong, C. Ko, J. Suh, K. Liu, K. Wang, J. J. Urban, and X. Zhang, 2017a, "Anomalously low electronic thermal conductivity in metallic vanadium dioxide," Science **355**, 371.





Lee, S., F. Yang, J. Suh, S. Yang, Y. Lee, G. Li, H. S. Choe, A. Suslu, Y. Chen, and C. Ko, 2015b, "Anisotropic in-plane thermal conductivity of black phosphorus nanoribbons at temperatures higher than 100 [thinsp] K," Nature Communications **6**, 8573.

Lepri, S., 1998, "Relaxation of classical many-body Hamiltonians in one dimension," Physical Review E **58**, 7165.

Lepri, S., R. Livi, and A. Politi, 1998, "On the anomalous thermal conductivity of one-dimensional lattices," EPL (Europhysics Letters) **43**, 271.

Lepri, S., R. Livi, and A. Politi, 2003, "Thermal conduction in classical low-dimensional lattices," Physics Reports **377**, 1.

Lepri, S., R. Livi, and A. Politi, "Thermal Transport in Low Dimensions," Lecture Notes in Physics Vol. 921 (Springer, Berlin, 2016), pp. 1–37

Li, H., H. Ying, X. Chen, D. L. Nika, A. I. Cocemasov, W. Cai, A. A. Balandin, and S. Chen, 2014, "Thermal conductivity of twisted bilayer graphene," Nanoscale **6**, 13402.

Li, H., Q. Zhang, C. C. R. Yap, B. K. Tay, T. H. T. Edwin, A. Olivier, and D. Baillargeat, 2012a, "From bulk to monolayer MoS2: evolution of Raman scattering," Advanced Functional Materials **22**, 1385.

Li, Q.-Y., K. Takahashi, H. Ago, X. Zhang, T. Ikuta, T. Nishiyama, and K. Kawahara, 2015, "Temperature dependent thermal conductivity of a suspended submicron graphene ribbon," Journal of Applied Physics **117**, 065102.

Li, T., 2012, "Ideal strength and phonon instability in single-layer MoS 2," Physical Review B **85**, 235407.



Li, W., L. Lindsay, D. Broido, D. A. Stewart, and N. Mingo, 2012b, "Thermal conductivity of bulk and nanowire Mg 2 Si x Sn 1− x alloys from first principles," Physical Review B **86**, 174307.

Li, W., N. Mingo, L. Lindsay, D. A. Broido, D. A. Stewart, and N. A. Katcho, 2012c, "Thermal conductivity of diamond nanowires from first principles," Physical Review B **85**, 195436.

Li, W., J. Carrete, and Natalio Mingo, 2013, "Thermal conductivity and phonon linewidths of monolayer MoS2 from first principles," Applied Physics Letters **103**, 253103.

Li, X., K. Maute, M. L. Dunn, and R. Yang, 2010, "Strain effects on the thermal conductivity of nanostructures," Physical Review B **81**, 245318.

Li, Z. Y, Y. Z Liu, L. Lindsay, Y. Xu, W-H. Duan, and E Pop, 2017, "Size Dependence and Ballistic Limits of Thermal Transport in Anisotropic Layered Two-Dimensional Materials," arXiv 1711.02772 .

Lindsay, L., and D. Broido, 2011a, "Enhanced thermal conductivity and isotope effect in single-layer hexagonal boron nitride," Physical Review B **84**, 155421.

Lindsay, L., and D. Broido, 2012, "Theory of thermal transport in multilayer hexagonal boron nitride and nanotubes," Physical Review B **85**, 035436.

Lindsay, L., D. Broido, and N. Mingo, 2010a, "Diameter dependence of carbon nanotube thermal conductivity and extension to the graphene limit," Physical Review B **82**, 161402.

Lindsay, L., D. Broido, and N. Mingo, 2010b, "Flexural phonons and thermal transport in graphene," Physical Review B **82**, 115427.

Lindsay, L., D. Broido, and N. Mingo, 2011, "Flexural phonons and thermal transport in multilayer graphene and graphite," Physical Review B **83**, 235428.





Lindsay, L., and D. A. Broido, 2011b, "Enhanced thermal conductivity and isotope effect in single-layer hexagonal boron nitride," Physical Review B **84**, 155421.

Lindsay, L., W. Li, J. Carrete, N. Mingo, D. Broido, and T. Reinecke, 2014, "Phonon thermal transport in strained and unstrained graphene from first principles," Physical Review B **89**, 155426.

Lindsay, L. R., 2010, Ph.D. thesis (Boston College).

Lippi, A., and R. Livi, 2000, "Heat conduction in two-dimensional nonlinear lattices," Journal of Statistical Physics **100**, 1147.

Liu, B., J. A. Baimova, C. D. Reddy, S. V. Dmitriev, W. K. Law, X. Q. Feng, and K. Zhou, 2014a, "Interface thermal conductance and rectification in hybrid graphene/silicene monolayer," Carbon **79**, 236.

Liu, B., C. Reddy, J. Jiang, J. A. Baimova, S. V. Dmitriev, A. A. Nazarov, and K. Zhou, 2012a, "Morphology and in-plane thermal conductivity of hybrid graphene sheets," Applied Physics Letters **101**, 211909.

Liu, B., C. Reddy, J. Jiang, J. A. Baimova, S. V. Dmitriev, A. A. Nazarov, and K. Zhou, 2012b, "Morphology and in-plane thermal conductivity of hybrid graphene sheets," Applied Physics Letters **101**, 211909.

Liu, F., P. Ming, and J. Li, 2007, "Ab initio calculation of ideal strength and phonon instability of graphene under tension," Physical Review B **76**, 064120.

Liu, H., G. Qin, Y. Lin, and M. Hu, 2016, "Disparate Strain Dependent Thermal Conductivity of Two-dimensional Penta-Structures," Nano Letters **16**, 3831.





Liu, J., G.-M. Choi, and D. G. Cahill, 2014b, "Measurement of the anisotropic thermal conductivity of molybdenum disulfide by the time-resolved magneto-optic Kerr effect," Journal of Applied Physics **116**, 233107.

Liu, J., and R. Yang, 2012, "Length-dependent thermal conductivity of single extended polymer chains," Physical Review B **86**, 104307.

Liu, S., P Hänggi, N. Li, J. Ren, B. Li, 2014 "Anomalous heat diffusion," Physical Review Letters 112, 040601.

Liu, T.-H., Y.-C. Chen, C.-W. Pao, and C.-C. Chang, 2014c, "Anisotropic thermal conductivity of MoS2 nanoribbons: Chirality and edge effects," Applied Physics Letters **104**, 201909.

Liu, X., G. Zhang, Q.-X. Pei, and Y.-W. Zhang, 2013, "Phonon thermal conductivity of monolayer MoS2 sheet and nanoribbons," Applied Physics Letters **103**, 133113.

Luo, Z., J. Maassen, Y. Deng, Y. Du, R. P. Garrelts, M. S. Lundstrom, D. Y. Peide, and X. Xu, 2015, "Anisotropic in-plane thermal conductivity observed in few-layer black phosphorus," Nature Communications **6**, 8572.

Mai, T., and O. Narayan, 2006, "Universality of one-dimensional heat conductivity," Physical Review E **73**, 061202.

Majee, A. K. and Z. Aksamija, 2016, "Length divergence of the lattice thermal conductivity in suspended graphene nanoribbons," Physical Review B 93, 235423.

Mak, K. F., C. H. Lui, and T. F. Heinz, 2010, "Measurement of the thermal conductance of the graphene/SiO2SiO2 interface," Applied Physics Letters, **97**, 221904.

Malekpour, H., P. Ramnani,, S. Srinivasan, G. Balasubramanian, D. L. Nika, A. Mulchandani, R. K. Lake, and A. A. Balandin, 2016, "Thermal conductivity of graphene with defects induced by electron beam irradiation," Nanoscale **8**, 14608.



Mariani, E., and F. von Oppen, 2008, "Flexural phonons in free-standing graphene," Physical Review Letters **100**, 076801.

Maruyama, S., 2002, "A molecular dynamics simulation of heat conduction in finite length SWNTs," Physica B: Condensed Matter **323**, 193.

Mingo, N., and D. Broido, 2005, "Length dependence of carbon nanotube thermal conductivity and the "problem of long waves," Nano Letters **5**, 1221.

Mingo, N., K. Esfarjani, D. Broido, and D. Stewart, 2010, "Cluster scattering effects on phonon conduction in graphene," Physical Review B **81**, 045408.

Mohiuddin T. M. G. , A. Lombardo, R. R. Nair,  A. Bonetti, G. Savini, R. Jalil,  N. Bonini, D. M. Basko, C. Galiotis, N. Marzari, K. S. Novoselov,1 A. K. Geim, and A. C. Ferrari, 2009 "Uniaxial strain in graphene by Raman spectroscopy: G peak splitting, Grüneisen parameters, and sample orientation," Physical Review B **79**, 205433.

Morelli, D., and J. Heremans, 2002, "Thermal conductivity of germanium, silicon, and carbon nitrides," Applied Physics Letters **81**, 5126.

Mounet N., and N. Marzari 2005, "First-principles determination of the structural, vibrational and thermodynamic properties of diamond, graphite, and derivatives," Physical Review B **71**, 205214.

Mu, X., X. Wu, T. Zhang, D. B. Go, and T. Luo, 2014, "Thermal transport in graphene oxide–from ballistic extreme to amorphous limit," Scientific Reports **4**, 3909.

Narayan, O., and S. Ramaswamy, 2002, "Anomalous heat conduction in one-dimensional momentum-conserving systems," Physical Review Letters **89**, 200601.

Narayanamurti, V., and R. Dynes, 1972, "Observation of second sound in bismuth," Physical Review Letters **28**, 1461.





Ni, Y., Y. Chalopin, and S. Volz, 2013, "Few layer graphene based superlattices as efficient thermal insulators," Applied Physics Letters, **103**, 141905.

Nika, D. L., S. Ghosh, E. P. Pokatilov, and A. A. Balandin, 2009a, "Lattice thermal conductivity of graphene flakes: Comparison with bulk graphite," Applied Physics Letters, **94**, 203103.

Nika, D. L., E. P. Pokatilov, A. S. Askerov, and A. A. Balandin, 2009b, "Phonon thermal conduction in graphene: Role of Umklapp and edge roughness scattering," Applied Physics Letters, **79**, 155413.

Nika, D. L., A. S. Askerov, and A. A. Balandin, 2012, "Anomalous size dependence of the thermal conductivity of graphene ribbons," Nano Letters **12**, 3238.

Nika, D. L., and A. A. Balandin, 2012, "Two-dimensional phonon transport in graphene," Journal of Physics: Condensed Matter **24**, 233203.

Nika, D. L., and A. A. Balandin, 2017, "Phonons and thermal transport in graphene and graphene-based materials," Reports on Progress in Physics **80**, 036502.

Nika, D. L., A. I. Cocemasov, and A. A. Balandin, 2014, "Specific heat of twisted bilayer graphene: Engineering phonons by atomic plane rotations," Applied Physics Letters **105**, 031904.

Nguyen, V. H., M. C. Nguyen, H. V. Nguyen, J. Saint-Martin, and P. Dollfus, 2014, "Enhanced thermoelectric figure of merit in vertical graphene junctions," Applied Physics Letters, **105**, 133105.

Omini, M., and A. Sparavigna, 1996, "Beyond the isotropic-model approximation in the theory of thermal conductivity," Physical Review B **53**, 9064.

Ong, Z.-Y., and E. Pop, 2011, "Effect of substrate modes on thermal transport in supported graphene," Physical Review B **84**, 075471.





Ouyang, T., Y. Chen, L.-M. Liu, Y. Xie, X. Wei, and J. Zhong, 2012, "Thermal transport in graphyne nanoribbons," Physical Review B **85**, 235436.

Ouyang, T., Y. Chen, Y. Xie, K. Yang, Z. Bao, and J. Zhong, 2010, "Thermal transport in hexagonal boron nitride nanoribbons," Nanotechnology **21**, 245701.

Palomaki, T., J. Harlow, J. Teufel, R. Simmonds, and K. Lehnert, 2013, "Coherent state transfer between itinerant microwave fields and a mechanical oscillator," Nature **495**, 210.

Park, M., S.-C. Lee, and Y.-S. Kim, 2013, "Length-dependent lattice thermal conductivity of graphene and its macroscopic limit," Journal of Applied Physics **114**, 053506.

Parrish, K. D., A. Jain, J. M. Larkin, W. A. Saidi, and A. J. McGaughey, 2014, "Origins of thermal conductivity changes in strained crystals," Physical Review B **90**, 235201.

Pei, Q.-X., Z.-D. Sha, and Y.-W. Zhang, 2011, "A theoretical analysis of the thermal conductivity of hydrogenated graphene," Carbon **49**, 4752.

Peimyoo, N., J. Shang, W. Yang, Y. Wang, C. Cong, and T. Yu, 2015, "Thermal conductivity determination of suspended mono-and bilayer WS2 by Raman spectroscopy," Nano Research **8**, 1210.

Peng, B., H. Zhang, H. Shao, Y. Xu, X. Zhang, and H. Zhu, 2016, "Low lattice thermal conductivity of stanene," Scientific Reports **6**, 20225.

Pereira, L. F. C., and D. Donadio, 2013, "Divergence of the thermal conductivity in uniaxially strained graphene," Physical Review B **87**, 125424.

Pereverzev, A., 2003, "Fermi-Pasta-Ulam β lattice: Peierls equation and anomalous heat conductivity," Physical Review E **68**, 056124.

Pettes, M. T., I. Jo, Z. Yao, and L. Shi, 2011, "Influence of polymeric residue on the thermal conductivity of suspended bilayer graphene," Nano Letters **11**, 1195.





Pettes, M. T., J. Maassen, I. Jo, M. S. Lundstrom, and L. Shi, 2013, "Effects of surface band bending and scattering on thermoelectric transport in suspended bismuth telluride nanoplates," Nano Letters **13**, 5316.

Pettes, M. T., M. R. Sadeghi, H. Ji, I. Jo, W. Wei, R. S. Ruoff, and L. Shi, 2015, "Scattering of phonons by high-concentration isotopic impurities in ultrathin graphite," Physical Review B, **91**, 035429.

Picu, R., T. Borca-Tasciuc, and M. Pavel, 2003, "Strain and size effects on heat transport in nanostructures," Journal of Applied Physics **93**, 3535.

Pohl, D. W., and V. Irniger, 1976, "Observation of second sound in NaF by means of light scattering," Physical Review Letters **36**, 480.

Pop, E., V. Varshney, and A. K. Roy, 2012, "Thermal properties of graphene: Fundamentals and applications," MRS Bulletin **37**, 1273.

Poudel, B., Q. Hao, Y. Ma, Y. Lan, A. Minnich, B. Yu, X. Yan, D. Wang, A. Muto, and D. Vashaee, 2008, "High-thermoelectric performance of nanostructured bismuth antimony telluride bulk alloys," Science **320**, 634.

Prosen, T., and D. K. Campbell, 2000, "Momentum conservation implies anomalous energy transport in 1D classical lattices," Physical Review Letters **84**, 2857. It should be noted that later development found that momentum conservaton does not necessary mean an anomalous energy transport. For example, in the rotator model (Gendelman and Savin 2000, Giardina et al. 2000), although the (angular) momentum is conserved, the system demonstrtaes a fintite thermal conductivity. However, further analysis found that this model does not support a nonvanishing pressure, and thus the infinite-wavelength phonons cannot carry any energy current. Some recent numerical works also showed that




in some momentum conserved 1D system with asymmetric interparticle potential, the thermal conductivity is finite (Zhong et al. 2012, Chen et al. 2016). However, it was later demonstrated that this seemingly convergent thermal conductivity is a finite size effect. When the system size becomes larger, the system shows again a divergent thermal conductivity (Wang et al. 2013, Savin and Kosevich 2014, Das et al. 2014). More recent works by Gendelman and Savin 2014, and Sato (2016) showed that in 1D momentum conserved systems, thermally activated dissociation and increased pressure can make the thermal transport normal.


Qian, X., X. Gu, M. S. Dresselhaus, and R.G. Yang, 2016, "Anisotropic Tuning on Graphite Thermal Conductivity by Lithium Intercalation," The Journal of Physical Chemistry Letters **7**, 4744.

Qian, X., P.Q. Jiang, Y. Peng, X.K. Gu, Z. Liu, and R.G. Yang, 2018, "Anistropic Thermal Transport in van der Waals Layered Alloys WSe$_{2(1-x)}$Te$_{2x}$," Applied Physics Letters **112**, 241901.

Qin, G., and M. Hu, 2016, in *Diverse Thermal Transport Properties of Two-Dimensional Materials: A Comparative Review,* edited by Nayak, P. K. (InTech, Rijeka, Croatia).

Qin, G., Z. Qin, W.-Z. Fang, L.-C. Zhang, S.-Y. Yue, Z.-B. Yan, M. Hu, and G. Su, 2016b, " Diverse anisotropy of phonon transport in two-dimensional IV-VI compounds: A comparative study," Nanoscale **8**, 11306.

Qin, G., X. Zhang, S.-Y. Yue, Z. Qin, H. Wang, Y. Han, and M. Hu, 2016b, "Resonant bonding driven giant phonon anharmonicity and low thermal conductivity of phosphorene," Physical Review B **94**, 165445.





Qin, T., J. Zhou, and J. Shi, 2012, "Berry curvature and the phonon Hall effect," Physical Review B **86**, 104305.

Qiu, B., and X. Ruan, 2010, "Thermal conductivity prediction and analysis of few-quintuple Bi2Te3 thin films: A molecular dynamics study," Applied Physics Letters **97**, 183107.

Qiu, B., and X. Ruan, 2012, "Reduction of spectral phonon relaxation times from suspended to supported graphene," Applied Physics Letters **100**, 193101.

Ratsifaritana, C. A., and P. G. Klemens, 1987, "Scattering of phonons by vacancies," International Journal of Thermophysics **8**, 737.

Fleurial, J. P., L. Gailliard, R. Triboulet, H. Scherrer and S. Scherrer, 1988, "Thermal properties of high quality single crystals of bismuth telluride—Part I: Experimental characterization," Journal of Physics and Chemistry of Solids **49**, 1237.

Ross, R. G., P. Andersson, B. Sundqvist, and G. Backstrom, 1984, "Thermal conductivity of solids and liquids under pressure," Reports on Progress in Physics **47**, 1347.

Sadeghi, M. M., I. Jo, and L. Shi, 2013, "Phonon-interface scattering in multilayer graphene on an amorphous support," Proceedings of the National Academy of Sciences **110**, 16321.

Sadeghi, M. M., M. T. Pettes, and L. Shi, 2012, "Thermal transport in graphene," Solid State Communications **152**, 1321.

Sahoo, S., A. P. Gaur, M. Ahmadi, M. J.-F. Guinel, and R. S. Katiyar, 2013, "Temperature-dependent Raman studies and thermal conductivity of few-layer $MoS_2$," Journal of Physical Chemistry C **117**, 9042.

Sato, D. S., 2016, "Pressure-induced recovery of Fourier's law in one-dimensional momentum-conserving systems," Physical Review E **94**, 012115.





Savin, V., and Y. A. Kosevich, 2014, "Thermal conductivity of molecular chains with asymmetric potentials of pair interactions," Physical Review E **89**, 032102.

Seol, J. H., I. Jo, A. L. Moore, L. Lindsay, Z. H. Aitken, M. T. Pettes, X. Li, Z. Yao, R. Huang, and D. Broido, N. Mingo, R. S. Ruoff, and L. Shi, 2010, "Two-dimensional phonon transport in supported graphene," Science **328**, 213.

Sevik, C., A. Kinaci, J. B. Haskins, and T. Çağın, 2011, "Characterization of thermal transport in low-dimensional boron nitride nanostructures," Physical Review B **84**, 085409.

Sevinçli, H., W. Li, N. Mingo, G. Cuniberti, and S. Roche, 2011, "Effects of domains in phonon conduction through hybrid boron nitride and graphene sheets," Physical Review B **84**, 205444.

Shi, L., D. Li, C. Yu, W. Jang, D. Kim, Z. Yao, P. Kim, and A. Majumdar, 2003, "Measuring thermal and thermoelectric properties of one-dimensional nanostructures using a microfabricated device," Journal of Heat Transfer **125**, 881.

Singh, D., J. Y. Murthy, and T. S. Fisher, 2011, "Mechanism of thermal conductivity reduction in few-layer graphene," Journal of Applied Physics **110**, 044317.

Singh, V., S. Bosman, B. Schneider, Y. M. Blanter, A. Castellanos-Gomez, and G. Steele, 2014, "Optomechanical coupling between a multilayer graphene mechanical resonator and a superconducting microwave cavity," Nature Nanotechnology **9**, 820.

Slack, G. A., 1973, "Nonmetallic crystals with high thermal conductivity," Journal of Physics and Chemistry of Solids **34**, 321.

Song, J., and N. V. Medhekar, 2013, "Thermal transport in lattice-constrained 2D hybrid graphene heterostructures," Journal of Physics: Condensed Matter **25**, 445007.





Strohm, C., G. Rikken, and P. Wyder, 2005, "Phenomenological evidence for the phonon Hall effect," Physical Review Letters **95**, 155901.

Sullivan, S., A. Vallabhaneni, I. Kholmanov, X. Ruan, J. Murthy, and L. Shi, 2017, "Optical generation and detection of local non-equilibrium phonons in suspended graphene," Nano Letters **17**, 2049.

Sun, B., X. Gu, Q. Zeng, X. Huang, Y. Yan, Z. Liu, R. Yang, and Y. K. Koh, 2017, "Temperature Dependence of Anisotropic Thermal-Conductivity Tensor of Bulk Black Phosphorus," Advanced Materials **29**, 1603297.

Tamura, S.-i., 1983, "Isotope scattering of dispersive phonons in Ge," Physical Review B **27**, 858.

Tan, Z. W., J.-S. Wang, and C. K. Gan, 2010, "First-principles study of heat transport properties of graphene nanoribbons," Nano Letters **11**, 214.

Tang, X., S. Xu, J. Zhang, and X. Wang, 2014, "Five orders of magnitude reduction in energy coupling across corrugated graphene/substrate interfaces," ACS Applied Materials & Interfaces **6**, 2809.

Taube, A., J. Judek, A. Łapińska, and M. Zdrojek, 2015, "Temperature-dependent thermal properties of supported MoS2 monolayers," ACS Applied Materials & Interfaces **7**, 5061.

Teweldebrhan, D., V. Goyal, and A. A. Balandin, 2010, "Exfoliation and characterization of bismuth telluride atomic quintuples and quasi-two-dimensional crystals," Nano Letters **10**, 1209.

Tian, L., 2015, "Optoelectromechanical transducer: Reversible conversion between microwave and optical photons," Annalen der Physik **527**, 1.





Tran, V. T., J. Saint-Martin, P. Dollfus, and S. Volz, 2017, "Optimizing the thermoelectric performance of graphene nano-ribbons without degrading the electronic properties," Scientific Reports, **7**, 2313.

Vallabhaneni, A. K., D. Singh, H. Bao, J. Murthy, and X. Ruan, 2016, "Reliability of Raman measurements of thermal conductivity of single-layer graphene due to selective electron-phonon coupling: A first-principles study," Physical Review B **93**, 125432.

Venkatasubramanian, R., E. Siivola, T. Colpitts, and B. O'quinn, 2001, "Thin-film thermoelectric devices with high room-temperature figures of merit," Nature **413**, 597.

Wan, C., X. Gu, F. Dang, T. Itoh, Y. Wang, H. Sasaki, M. Kondo, K. Koga, K. Yabuki, and G. J. Snyder, 2015a, "Flexible n-type thermoelectric materials by organic intercalation of layered transition metal dichalcogenide TiS2," Nature Materials **14**, 622.

Wan, C., Y. Kodama, M. Kondo, R. Sasai, X. Qian, X. Gu, K. Koga, K. Yabuki, R. Yang, and K. Koumoto, 2015b, "Dielectric mismatch mediates carrier mobility in organic-intercalated layered TiS2," Nano Letters **15**, 6302.

Wan, C., Y. Wang, W. Norimatsu, M. Kusunoki, and K. Koumoto, 2012, "Nanoscale stacking faults induced low thermal conductivity in thermoelectric layered metal sulfides," Applied Physics Letters **100**, 101913.

Wan, C., Y. Wang, N. Wang, and K. Koumoto, 2010a, "Low-thermal-conductivity $(MS)_{1+x}$ $(TiS_2)_2$ (M= Pb, Bi, Sn) misfit layer compounds for bulk thermoelectric materials," Materials **3**, 2606.

Wan, C., Y. Wang, N. Wang, W. Norimatsu, M. Kusunoki, and K. Koumoto, 2010b, "Development of novel thermoelectric materials by reduction of lattice thermal conductivity," Science and Technology of Advanced Materials **11**, 044306.





Wan, C., Y. Wang, N. Wang, W. Norimatsu, M. Kusunoki, and K. Koumoto, 2011, "Intercalation: building a natural superlattice for better thermoelectric performance in layered Chalcogenides," Journal of Electronic Materials **40**, 1271.

Wang, F. Q., S. Zhang, J. Yu, and Q. Wang, 2015, "Thermoelectric properties of single-layered SnSe sheet," Nanoscale **7**, 15962.

Wang, H., and M. S. Daw, 2016, "Anharmonic renormalization of the dispersion of flexural modes in graphene using atomistic calculations," Physical Review B **94**, 155434.

Wang, J.Y., 2013, Ph.D. thesis (National Unviersity of Singapore).

Wang, L., B. Hu, and B. Li, 2012a, "Logarithmic divergent thermal conductivity in two-dimensional nonlinear lattices," Physical Review E **86**, 040101.

Wang, Li, B. Hu, and B. Li, 2013, "Validity of Fourier's law in one-dimensional momentum-conserving lattices with asymmetric interparticle interactions," Physical Review E **88**, 052112

Wang, Y., B. Qiu, and X. Ruan, 2012b, "Edge effect on thermal transport in graphene nanoribbons: A phonon localization mechanism beyond edge roughness scattering," Applied Physics Letters **101**, 013101.

Wang, Y., A. K. Vallabhaneni, B. Qiu, and X. Ruan, 2014, "Two-dimensional thermal transport in graphene: a review of numerical modeling studies," Nanoscale and Microscale Thermophysical Engineering **18**, 155.

Wang, Y., N. Xu, D. Li, and J. Zhu, 2017, "Thermal Properties of Two Dimensional Layered Materials," Advanced Functional Materials **27**, 1604134.

Ward, A., D. Broido, D. A. Stewart, and G. Deinzer, 2009, "Ab initio theory of the lattice thermal conductivity in diamond," Physical Review B **80**, 125203.





Wei, N., L. Xu, H.-Q. Wang, and J.-C. Zheng, 2011a, "Strain engineering of thermal conductivity in graphene sheets and nanoribbons: a demonstration of magic flexibility," Nanotechnology **22**, 105705.

Wei, Y., B. Wang, J. Wu, R. Yang, and M. L. Dunn, 2012a, "Bending rigidity and Gaussian bending stiffness of single-layered graphene," Nano Letters **13**, 26.

Wei, Z., Y. Chen, and C. Dames, 2012b, "Wave packet simulations of phonon boundary scattering at graphene edges," Journal of Applied Physics **112**, 024328.

Wei, Z., Z. Ni, K. Bi, M. Chen, and Y. Chen, 2011b, "In-plane lattice thermal conductivities of multilayer graphene films," Carbon **49**, 2653.

Wei, Z., J. Yang, K. Bi, and Y. Chen, 2014a, "Mode dependent lattice thermal conductivity of single layer graphene," Journal of Applied Physics **116**, 153503.

Wei, Z., J. Yang, W. Chen, K. Bi, D. Li, and Y. Chen, 2014b, "Phonon mean free path of graphite along the c-axis," Applied Physics Letters **104**, 081903.

Whittingha, S. M., 2012, *Intercalation chemistry* (Elsevier).

Wilson, J., and A. Yoffe, 1969, "The transition metal dichalcogenides discussion and interpretation of the observed optical, electrical and structural properties," Advances in Physics **18**, 193.

Xie, G., Y. Shen, X. Wei, L. Yang, H. Xiao, J. Zhong, and G. Zhang, 2014, "A Bond-order Theory on the Phonon Scattering by Vacancies in Two-dimensional Materials," Scientific Reports **4**, 5085.

Xie, H., L. Chen, W. Yu, and B. Wang, 2013, "Temperature dependent thermal conductivity of a free-standing graphene nanoribbon," Applied Physics Letters **102**, 111911.




Xie, H., T. Ouyang, É. Germaneau, G. Qin, M. Hu, and H. Bao, 2016, "Large tunability of lattice thermal conductivity of monolayer silicene via mechanical strain," Physical Review B **93**, 075404.

Xiong, D., J. Wang, Y. Zhang, and H. Zhao, 2010, "Heat conduction in two-dimensional disk models," Physical Review E **82**, 030101.

Xu, X., J. Chen, and B. Li, 2016, "Phonon thermal conduction in novel 2D materials," Journal of Physics: Condensed Matter **28**, 483001.

Xu, X., L. F. Pereira, Y. Wang, J. Wu, K. Zhang, X. Zhao, S. Bae, C. T. Bui, R. Xie, and J. T. Thong, 2014a, "Length-dependent thermal conductivity in suspended single-layer graphene," Nature Communications **5**, 3689.

Xu, Y., X. Chen, B.-L. Gu, and W. Duan, 2009, "Intrinsic anisotropy of thermal conductance in graphene nanoribbons," Applied Physics Letters **95**, 233116.

Xu, Y., and G. Li, 2009, "Strain effect analysis on phonon thermal conductivity of two-dimensional nanocomposites," Journal of Applied Physics **106**, 114302.

Xu, Y., Z. Li, and W. Duan, 2014b, "Thermal and thermoelectric properties of graphene," Small **10**, 2182.

Xu, Z., and M. J. Buehler, 2009, "Strain controlled thermomutability of single-walled carbon nanotubes," Nanotechnology **20**, 185701.

Yan, R., J. R. Simpson, S. Bertolazzi, J. Brivio, M. Watson, X. Wu, A. Kis, T. Luo, A. R. Hight Walker, and H. G. Xing, 2014, "Thermal conductivity of monolayer molybdenum disulfide obtained from temperature-dependent Raman spectroscopy," ACS Nano **8**, 986.





Yan. X.-J., Y.-Y. Lv, L. Li, X. Li, S.-H. Yao, Y.-B. Chen, X.-P. Liu, H. Lu, M.-H. Lu, and Y.-F. Chen, 2017, "Composition dependent phase transition and its induced hysteretic effect in the thermal conductivity of $W_xMo_{1-x}Te_2$," Applied Physics Letters **110**, 211904.

Yan, Z., C. Jiang, T. Pope, C. Tsang, J. Stickney, P. Goli, J. Renteria, T. Salguero, and A. Balandin, 2013, "Phonon and thermal properties of exfoliated $TaSe_2$ thin films," Journal of Applied Physics **114**, 204301.

Yang, J., E. Ziade, C. Maragliano, R. Crowder, X. Wang, M. Stefancich, M. Chiesa, A. K. Swan, and A. J. Schmidt, 2014, "Thermal conductance imaging of graphene contacts," Journal of Applied Physics **116**, 023515.

Yang, L., P. Grassberger, and B. Hu, 2006, "Dimensional crossover of heat conduction in low dimensions," Physical Review E **74**, 062101.

Yang, N., X. Xu, G. Zhang, and B. Li, 2012, "Thermal transport in nanostructures," AIP Advances **2**, 041410.

Yang, N., G. Zhang, and B. Li, 2010, "Violation of Fourier's law and anomalous heat diffusion in silicon nanowires," Nano Today **5**, 85.

Yang, R., and G. Chen, 2004, "Thermal conductivity modeling of periodic two-dimensional nanocomposites," Physical Review B **69**, 195316.

Yang, R., G. Chen, and M. S. Dresselhaus, 2005, "Thermal conductivity of simple and tubular nanowire composites in the longitudinal direction," Physical Review B **72**, 125418.

Ye, Z.-Q., B.-Y. Cao, W.-J. Yao, T. Feng, and X. Ruan, 2015, "Spectral phonon thermal properties in graphene nanoribbons," Carbon **93**, 915.

Zabel, H., 2001, "Phonons in layered compounds," Journal of Physics: Condensed Matter **13**, 7679.





Zeraati, M., S. M. V. Allaei, I. A. Sarsari, M. Pourfath, and D. Donadio, 2016, "Highly anisotropic thermal conductivity of arsenene: An ab initio study," Physical Review B **93**, 085424.

Zhang, G., and B. Li, 2005, "Thermal conductivity of nanotubes revisited: Effects of chirality, isotope impurity, tube length, and temperature," The Journal of Chemical Physics **123**, 114714.

Zhang, G., and Y.-W. Zhang, 2017, "Thermal properties of two-dimensional materials," Chinese Physics B **26**, 034401.

Zhang, H., X. Chen, Y.-D. Jho, and A. J. Minnich, 2016, "Temperature-Dependent Mean Free Path Spectra of Thermal Phonons Along the c-Axis of Graphite," Nano Letters **16**, 1643.

Zhang, H., G. Lee, and K. Cho, 2011, "Thermal transport in graphene and effects of vacancy defects," Physical Review B **84**, 115460.

Zhang, H., C.-X. Liu, X.-L. Qi, X. Dai, Z. Fang, and S.-C. Zhang, 2009, "Topological insulators in Bi2Se3, Bi2Te3 and Sb2Te3 with a single Dirac cone on the surface," Nature Physics **5**, 438.

Zhang, J., Y. Hong, and Y. Yue, 2015a, "Thermal transport across graphene and single layer hexagonal boron nitride," Journal of Applied Physics **117**, 134307.

Zhang, L., J. Ren, J.-S. Wang, and B. Li, 2010, "Topological nature of the phonon Hall effect," Physical Review Letters **105**, 225901.

Zhang, X., D. Sun, Y. Li, G.-H. Lee, X. Cui, D. Chenet, Y. You, T. F. Heinz, and J. C. Hone, 2015b, "Measurement of lateral and interfacial thermal conductivity of single-and bilayer MoS2 and MoSe2 using refined optothermal raman technique," ACS Applied Materials & Interfaces **7**, 25923.





Zhang, Z.-W., S.-Q. Hu, J. Chen, and B. Li, 2017, "Hexagonal Boronnitride: A promising substrate for graphene with high heat dissipation," Nanotechonology **28**, 226704.

Zhong, W.-R., M.-P. Zhang, B.-Q. Ai, and D.-Q. Zheng, 2011, "Chirality and thickness-dependent thermal conductivity of few-layer graphene: a molecular dynamics study," Applied Physics Letters **98**, 113107.

Zhong Y., Y. Zhang, J. Wang, and H. Zhao, 2012 "Normal heat conduction in one-dimensional momentum conserving lattices with asymmetric interactions," Physical Review E **85**, 060102.

Zhu, G., J. Liu, Q. Zheng, R. Zhang, D. Li, D. Banerjee, and D. G. Cahill, 2016a, "Tuning thermal conductivity in molybdenum disulfide by electrochemical intercalation," Nature Communications **7**, 13211.

Zhu, J., H. Park, J. Y. Chen, X. Gu, H. Zhang, S. Karthikeyan, N. Wendel, S. A. Campbell, M. Dawber, and X. Du, 2016b, "Revealing the Origins of 3D Anisotropic Thermal Conductivities of Black Phosphorus," Advanced Electronic Materials **2**, 1600040.

Zhu, L., G. Zhang, and B. Li, 2014, "Coexistence of size-dependent and size-independent thermal conductivities in phosphorene," Physical Review B **90**, 214302.

Zhu, T., and E. Ertekin, 2014, "Phonon transport on two-dimensional graphene/boron nitride superlattices," Physical Review B **90**, 195209.

Ziman, J. M., 1960, *Electrons and phonons: the theory of transport phenomena in solids* (Oxford university press).




# Table

(Sahoo, *et al.*, 2013; Jo, *et al.*, 2014; Liu, *et al.*, 2014b; Yan, *et al.*, 2014; Zhang, *et al.*, 2015b),

**Table I.** Experimental studies on the thermal conductivity of MoS$_2$. χ is the temperature coefficient of Raman signal and α is the absorption ratio used for data fitting.

| Ref. | Method | Sample type | Room-temperature thermal conductivity (W/mK) | Experimental conditions |
|------|--------|-------------|-------------|-------------------------|
| Yan, *et al.*, 2014 | Raman | Exfoliated, transferred | 34.5±4 (1-Layer) | A$_{1g}$ mode, χ=0.011 cm$^{-1}$/K, α=9±1% ,170 nm diameter laser spot, suspended on 1.2-μm-diameter holes, ambient condition |
| Zhang, *et al.*, 2015b | Raman | Exfoliated, transferred | 84±17 (1-Layer) | A$_{1g}$ mode, χ =0.0203 cm$^{-1}$/K, α=5.2±0.1%,460-620 nm diameter laser spot, suspended on 2.5-to-5.0-μm-diameter holes, ambient condition |
| Zhang, *et al.*, 2015b | Raman | Exfoliated, transferred | 77±25 (2-Layer) | A$_{1g}$ mode, χ =0.0136 cm$^{-1}$/K, α=11.5±0.1%,460-620 nm diameter laser spot, suspended on 2.5-to-5.0-μm-diameter holes, ambient condition |
| Jo, *et al.*, 2014 | Micro-bridge | Exfoliated, transferred | 44–50 (4-Layer) | Suspended sample; length: 3 μm, width: 5.2 μm. |
| Jo, *et al.*, 2014 | Micro-bridge | Exfoliated, transferred | 48-52 (7-Layer) | Suspended sample; length: 8 μm, width: 2.2 μm. |
| Sahoo, *et al.*, 2013 | Raman | CVD, transferred | 52 (11-Layer) | A$_{1g}$ mode, χ= 1.23 × 10$^{-2}$cm/K, α = 10%,1-1.5 μm laser spot, suspended on a 10- μm-radius quadrant, ambient condition |
| Liu, *et al.*, 2014b | Pump-probe | Bulk | 85-112 | Modulation frequency of pump beam: 10.7 MHz |



# Figure Captions

**Figure 1.** Thermal conductivities of some typical single-layer 2-D materials. (a) graphene; (b) boron nitride, silicene, stanene, and blue phosphorene, which are of hexagonal crystal structures; (c) a few transition metal dichalcogenides; (d) a few 2-D crystals with puckered structures.

**Figure 2.** Comparison of thermal conductivities of some 2-D materials calculated from the Slack equation (x axis) and from the first-principle-based PBTE (y axis).

**Figure 3.** Thermal conductivities of some typical single-layer 2-D materials at room temperature. The data are sorted by descending thermal conductivity values. Inset: the atomic structures of boron nitride, silicene, $MoS_2$, and $ZrS_2$ as examples of 2-D materials with different crystal structures.

**Figure 4.** (a) Scanning electron microscopy image of the micro-bridge measurement device, which consists of two $Pt/SiN_x$ membranes. The red and blue Pt coils are the heater and temperature sensors, which are thermally connected by suspended graphene (grey sheet in the middle). Scale bar, 5 mm. (b) $SiN_x$ membrane-based heater structures optimized for length-dependent studies. Scale bar, 20 mm. Adapted from (Xu*, et al.*, 2014a) with permission from the Nature Publishing Group, Copyright 2014. (c) Thermal conductivity of graphene as a function of sample size. (d) Thermal conductivity of some 2-D materials as a function of the sample size.



**Figure 5.** (a) Schematic representation of phonon annihilation processes using the phonon dispersion curves of single layer graphene as an example. Red: the annihilation of a phonon mode generating a third phonon on the same branch as the second one; blue: the annihilation of a phonon mode generating a third phonon on a different branch from the second one. (b) Phonon decay processes. The solid circles are the phonon modes taking part in the scatterings. The solid circles pointed by arrows are the final states after scatterings, while others are the initial states before scattering.

**Figure 6.** Scattering rates for the ZA phonon modes along the $\Gamma-K$ direction in (a) unstrained graphene and (b) graphene under 0.5% tensile strain at 300 K. Inset: Contributions to the total rates: ZA + ZA → TA (LA) (open circles) and ZA + ZO → LA (TA) or ZA + LA (TA) → ZO (solid circles). Adapted from (Bonini, et al., 2012) with permission from the American Chemical Society, Copyright 2012.

**Figure 7.** Thickness-dependent thermal conductivity of (a) graphene, (b) $MoS_2$, (c) $Bi_2Te_3$, and (d) black phosphorene. The data in (d) comes from (1) (Lee, *et al.*, 2015b), (2) (Luo, *et al.*, 2015), (3) (Jang, *et al.*, 2015), (4) (Zhu, *et al.*, 2016b), (5)(Sun, *et al.*, 2017), (6) (Zhu, *et al.*, 2014), (7) (Jain and McGaughey, 2015), (8) (Qin, *et al.*, 2016b).

**Figure 8**. (a) Thermal conductivity of some 2-D materials as a function of tensile strain. (b) Phonon dispersion of silicene under different tensile strain. Adapted from (Xie, *et al.*, 2016) with the permission from the American Society of Sciences, Copyright 2016.



**Figure 9.** (a) Measured thermal conductivity of supported graphene (on $SiO_2$) together with the highest reported values of pyrolytic graphite. PG means pyrolytic graphite, and G1, G2, and G3 refer to three graphene samples. Adapted from (Seol*, et al.*, 2010) with the permission from the American Association for the Advancement of Science. (b) Measured room temperature thermal conductivity of supported graphene (on $SiN_x$) as a function of number of atomic layers (red star) with the best-fit curve (red line).  Inset: Temperature dependence of thermal conductivity around 300 K. (Courtesy: (Wang, 2013))

**Figure 10.** Thermal conductivity of supported or encased few-layer graphene at room temperature as a function of sample thickness.

**Figure 11.** (a) Basal-plane thermal conductivity and (b) thermal conductivity along the c-axis, as a function of lithium composition. (c-d) Comparison of phonon dispersion in the basal plane of pristine graphite and $LiC_6$. (e-f) Comparison of phonon dispersion along the c-axis of pristine graphite and $LiC_6$. Adapted from (Xin, *et. al.*, 2016) with the permission from American Chemical Society, Copyright 2016.

**Figure 12.** (a) High angle annular dark field scanning transmission electron microscopy image of $TiS_2[(HA)_x(H_2O)_y(DMSO)_z]$ along with the schematic atomic structure. (b) Measured in-plane thermal conductivity. Adapted from (Wan*, et al.*, 2015a) with the permission from the Nature Publishing Group, Copyright 2015.



**Figure 1**

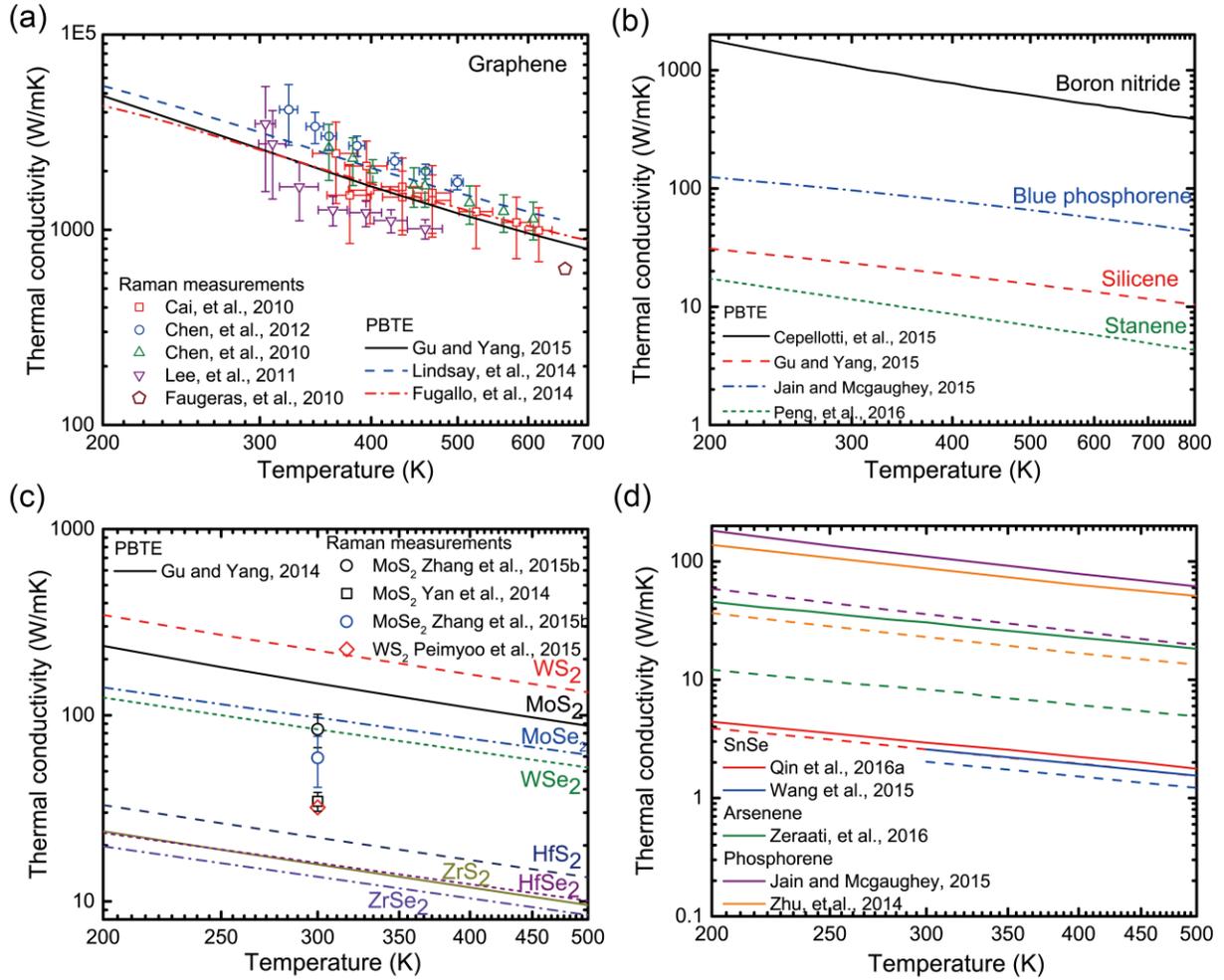



**Figure 2**

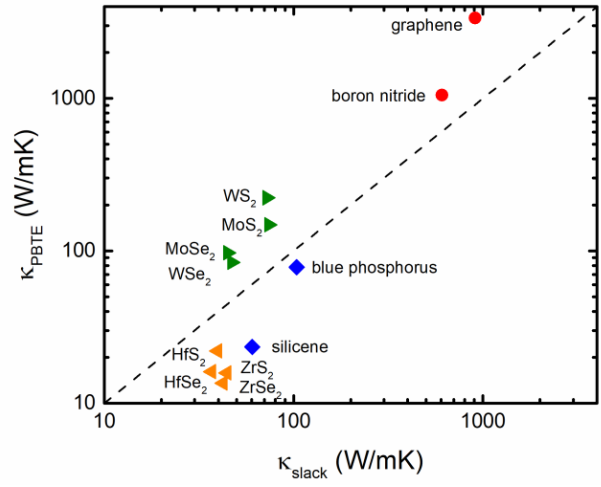





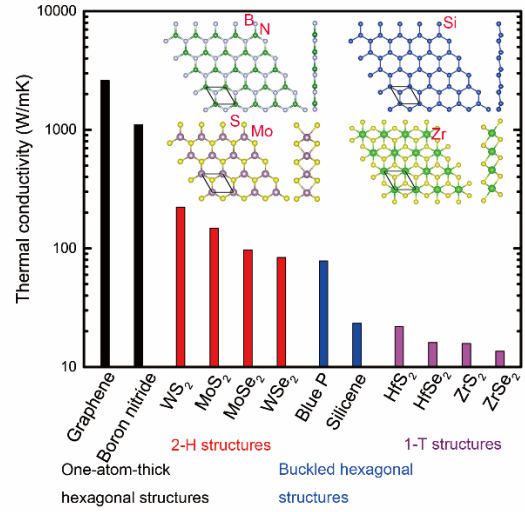



**Figure 4**

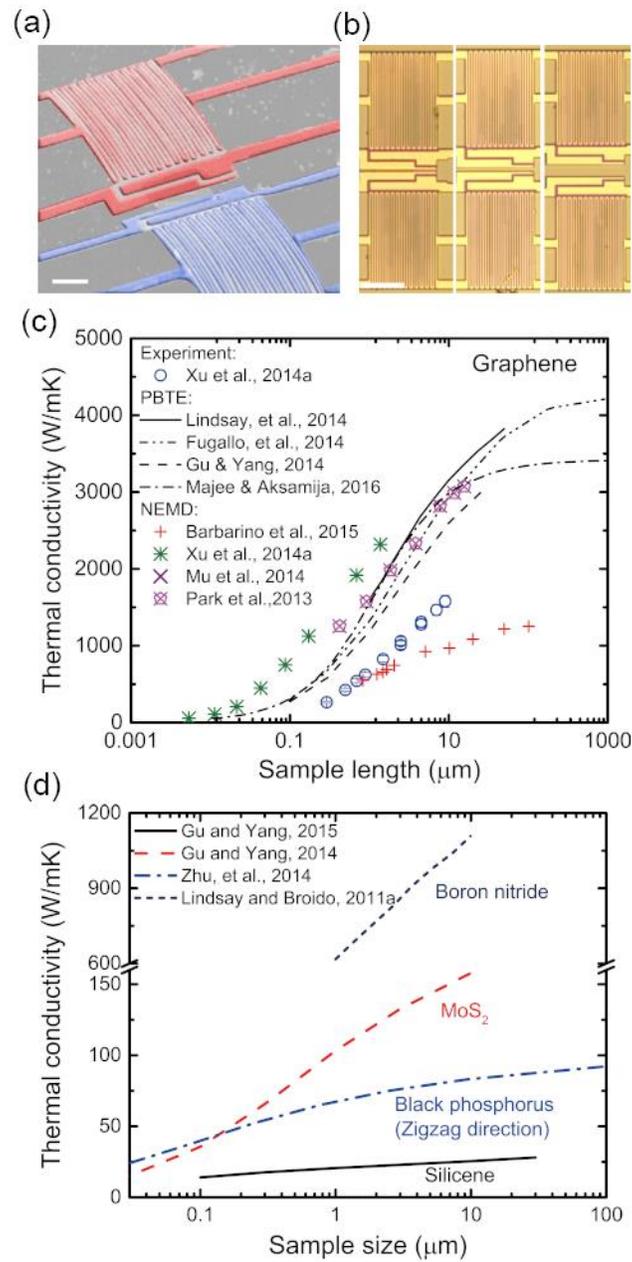

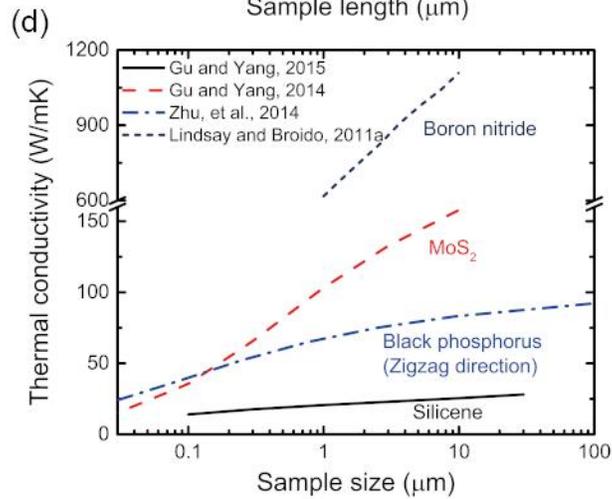



**Figure 5**

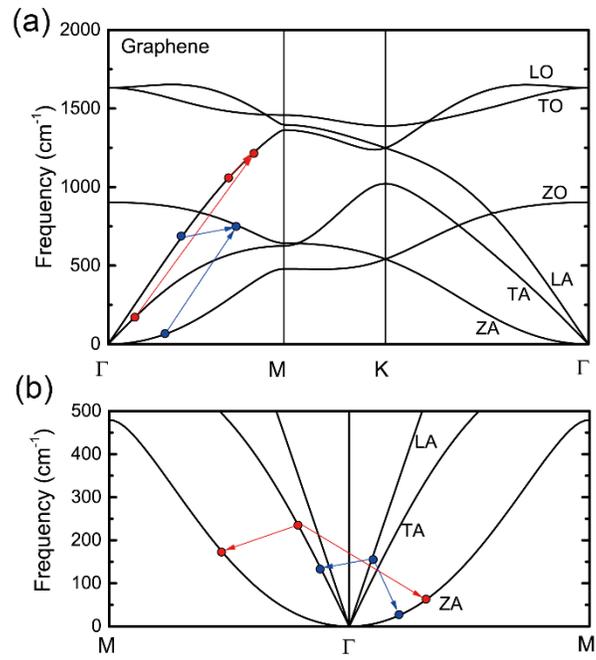





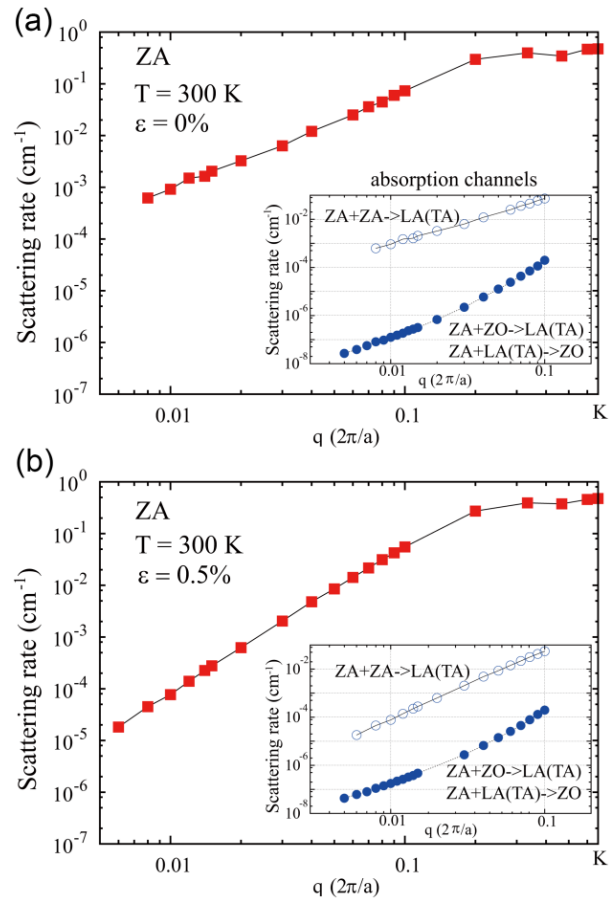



**Figure 7**

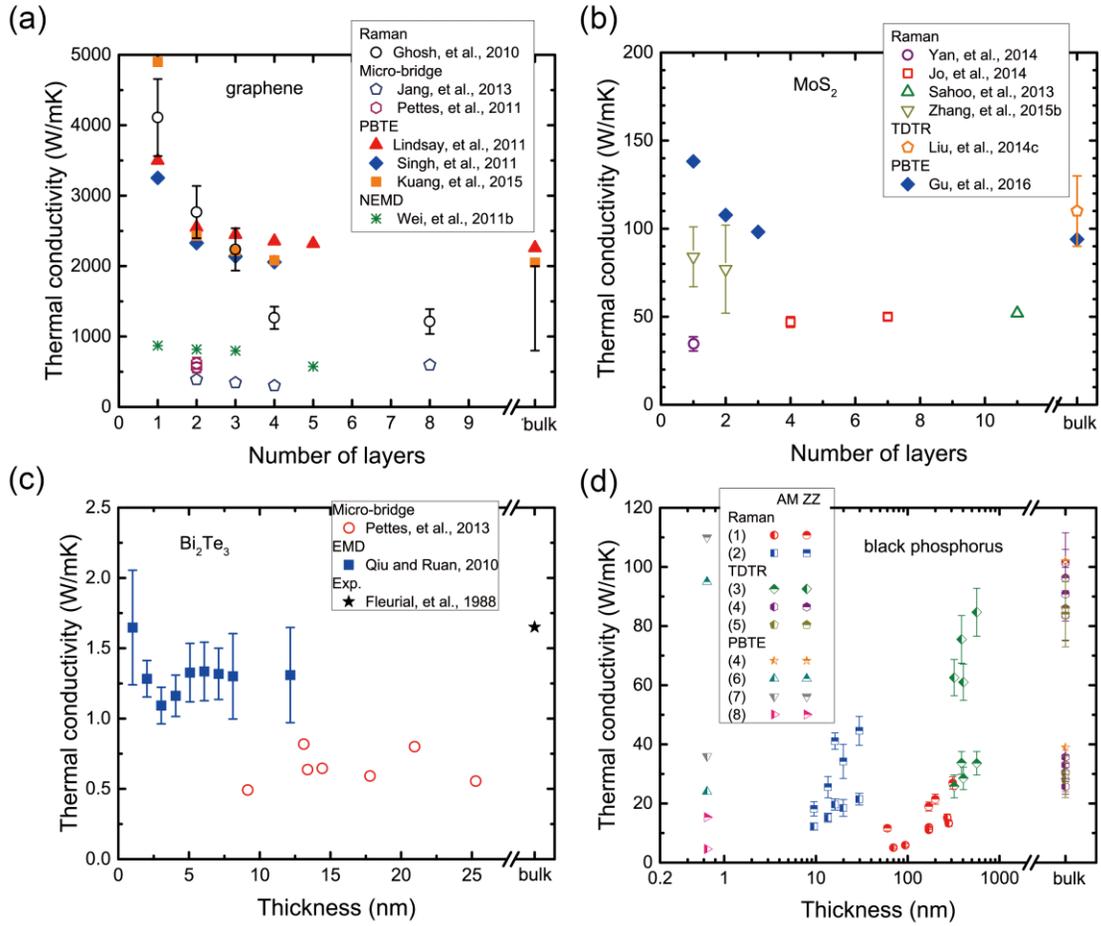



**Figure 8**

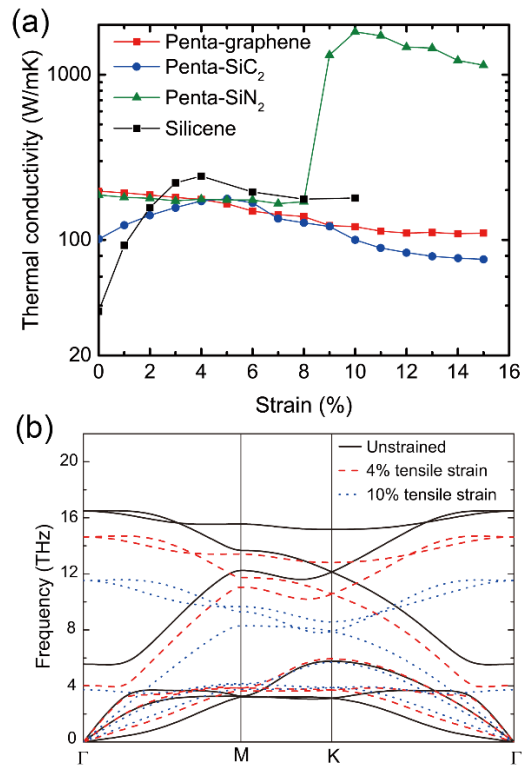



**Figure 9**

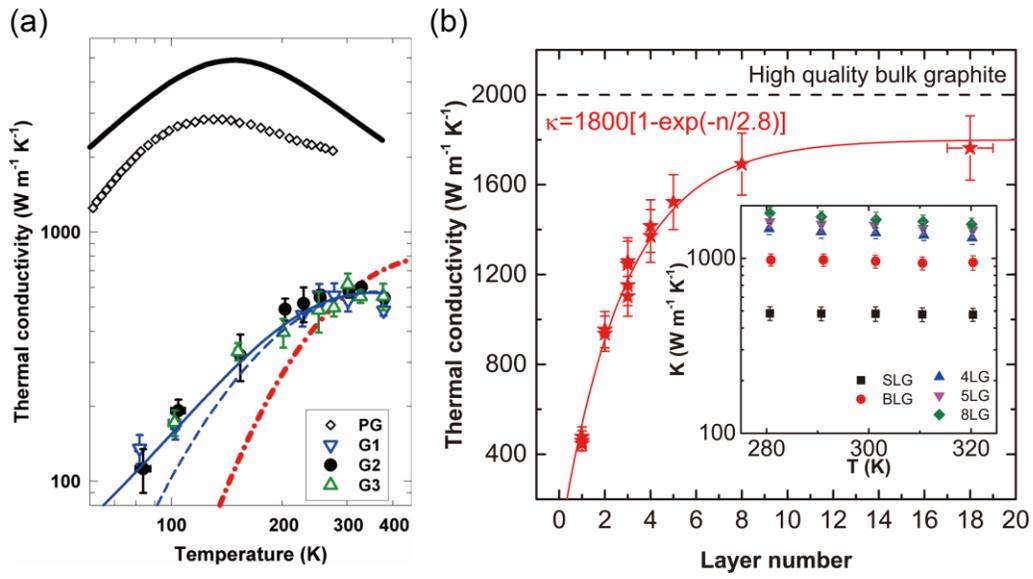



**Figure 10**

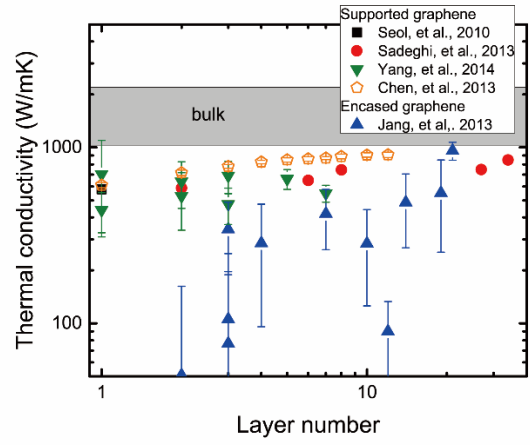





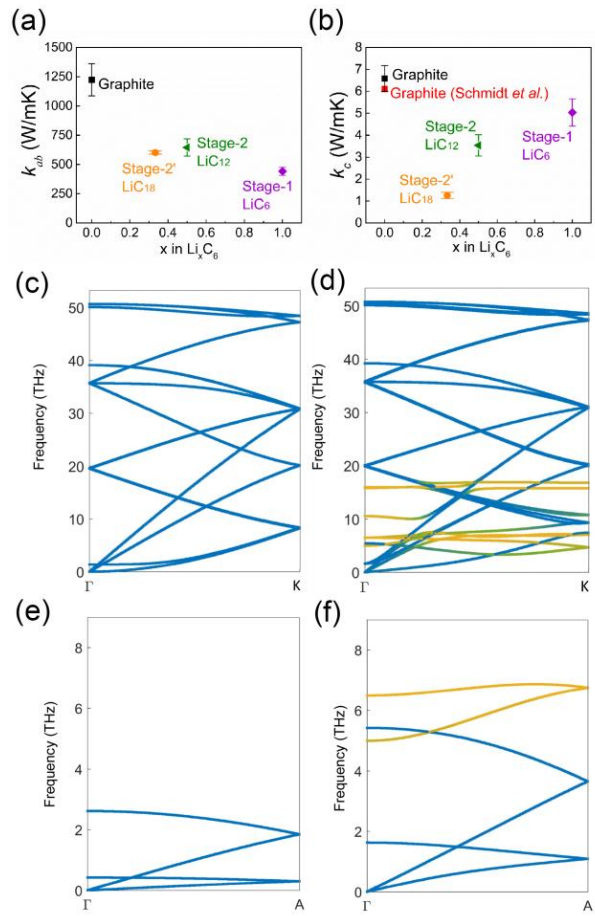



**Figure 12**

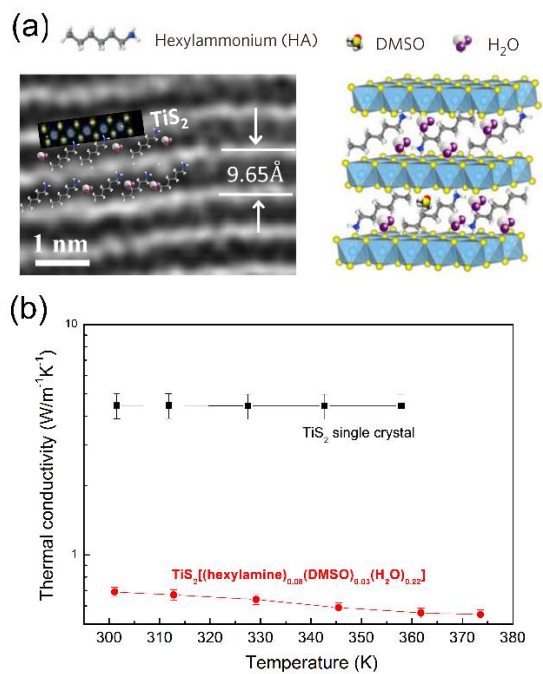